\documentclass[aip,
pof,
amsmath,amssymb,
preprint,
reprint,
]{revtex4-1}

\usepackage{graphicx}
\usepackage{dcolumn}
\usepackage{booktabs}
\usepackage{bm}

\usepackage[utf8]{inputenc}
\usepackage[T1]{fontenc}
\usepackage{mathptmx}
\usepackage{etoolbox}

\makeatletter
\def\@email#1#2{%
 \endgroup
 \patchcmd{\titleblock@produce}
  {\frontmatter@RRAPformat}
  {\frontmatter@RRAPformat{\produce@RRAP{*#1\href{mailto:#2}{#2}}}\frontmatter@RRAPformat}
  {}{}
}%
\makeatother

\begin{document}

\preprint{AIP/123-QED}

\title{Roughness-controlled layer in oscillatory turbulent boundary layers over densely packed uniform roughness}
\author{Xuchen Liu}
\author{Yuan Gao}
\author{Yiyong Dong}
\author{Jing Yuan}
\affiliation{Department of Hydraulic Engineering, Tsinghua University, Beijing, 100084, China}
\email{yuanj2021@mail.tsinghua.edu.cn}

\date{\today}

\begin{abstract}

In coastal wave boundary layers over gravel-scale roughness, with near-bed orbital excursions ten to a hundred times the roughness height, the boundary layer is only a few roughness heights thick. A roughness-controlled layer (RCL) of the steady-flow extent two to five element heights would then leave no room for a logarithmic layer, yet experiments over densely packed marbles recover logarithmic profiles within millimetres of the crests. We resolve this contradiction by re-analysing previous Particle Image Velocimetry (PIV) records with a triple decomposition that separates the marble-locked dispersive motion from the stochastic turbulence, across eleven wave, current and wave-current conditions. The boundary layer organises into an RCL, a transition region, and a logarithmic profile layer, with the RCL only one to two tenths of a marble diameter deep. This thinness is kinematic: above a periodic bed, the dispersive field decays over a length fixed by the element spacing, so close packing caps the layer at a fraction of a diameter. The layer is destroyed and rebuilt every half-cycle, tracking the near-bed velocity quasi-steadily, while the eddies within it stay locked to the inter-crest gap. Thinness does not imply weakness: within the layer, the dispersive kinetic energy rivals the turbulent kinetic energy, and the dispersive stress matches, near the crests exceeds, the Reynolds stress, showing that separated wakes carry organised momentum. The logarithmic layer survives because the decay length imposed by the packing is far smaller than the boundary-layer thickness, a margin set by the bed geometry rather than by the forcing.

\end{abstract}


\maketitle

\section{Introduction}\label{sec:intro}

In coastal waters, surface waves and currents interact within a thin bottom boundary layer that controls the bottom shear stress, the vertical mixing, and ultimately, the sediment transport \cite{Grant1979, Nielsen1992, Madsen1994}. Under field conditions capable of moving sediment, the near-bottom excursion amplitude $A_{bm}$ reaches $O(1\ \mathrm{m})$ and the orbital velocity $U_{orb}$ reaches $O(1\ \mathrm{m/s})$, so the amplitude Reynolds number $Re = A_{bm}U_{orb}/\nu$, with $\nu$ the kinematic viscosity, is of $O(10^6)$ and the boundary layer is fully turbulent \cite{Sleath1987}. The layer is nonetheless thin, and its thickness $\delta_w$ scales with $l_w = \kappa u_{*w}/\omega$, where $\kappa$ is the von Kármán constant, $u_{*w}$ the wave shear velocity and $\omega$ the angular frequency, and is only $O(1$--$10\ \mathrm{cm})$. Over a sand bed, whose Nikuradse roughness $k_N$ is $O(1\ \mathrm{mm})$, the governing length scales are therefore strongly ordered,

\begin{equation}
A_{bm} \gg \delta_w \gg k_N. 
\label{hierarchy}
\end{equation}
This hierarchy is the structural premise of the classical description of turbulent wave-current boundary layers. The upper separation, $A_{bm} \gg \delta_w$, places the flow in the fully developed turbulent regime characterised by the relative roughness $A_{bm}/k_N$ \cite{Sleath1987, Madsen1994}; the lower separation, $\delta_w \gg k_N$, leaves room between the roughness and the top of the wave boundary layer for an overlap region in which the velocity profile can be logarithmic, in direct analogy with the inertial sublayer of steady wall turbulence. Theoretical models of the Grant–Madsen group \cite{Grant1979, Madsen1994, GRM2011} are built on this logarithmic layer, and their predictions of bottom shear stress and apparent roughness inherit its validity.

The logarithmic profile in oscillatory flow also has a dynamical justification. Near the velocity extrema, the free-stream acceleration and hence the driving pressure gradient nearly vanish, so the near-bottom flow is quasi-steady and momentarily obeys the same wall-bounded dynamics as a steady flow \cite{Grant1977, TM1984a, Yuan2014}. Instantaneous log-profile fitting fails only in short windows around flow reversal, roughly $20$--$30\%$ of the wave period \cite{Yuan2014}. The price of this quasi-steady validity is spatial: the logarithmic region is confined to about $0.1l_w$ from the bottom, within $O(1\ \mathrm{mm})$--$O(10\ \mathrm{mm})$ of the bed — and therefore to within a few roughness heights whenever $k_N$ is not vanishingly small.

Because the required values of $Re$ and $A_{bm}/k_N$ are unattainable in small-scale facilities, the experimental basis of this picture comes almost entirely from full-scale oscillating water tunnels (OWTs), and the accumulated evidence divides sharply by relative roughness. For sand-scale roughness, $A_{bm}/k_N \gtrsim O(10^{2\text{-}3})$, the logarithmic layer is robust. Measurements over sand and sandpaper beds, spanning sinusoidal, Stokes, forward-leaning and acceleration-skewed waves \cite{Sleath1987, Jensen1989, vanderA2011, Yuan2014}, recover near-bottom logarithmic profiles accurately enough that log-profile fitting yields the theoretical bed level and bottom roughness, and provides the only reliable estimate of the bottom shear stress \cite{Yuan2014}. The turbulence intensity and Reynolds-stress structure in this regime are likewise documented \cite{Sleath1987, Jensen1989, vanderA2011, Hay2012a, Hay2012b, Yuan2015}. At the opposite extreme, $A_{bm}/k_N \sim O(1)$, sits the vortex-ripple bed, whose hydraulic roughness is set by the ripple height $\eta$ of $O(1$--$10\ \mathrm{cm})$: the scale separation collapses ($\delta_w \sim \eta$), spanwise coherent vortices shed from the crests dominate the entire layer, and no logarithmic region survives\cite{Fredsoe1999, vanderWerf2007, OnderYuan2019, YuanWang2019, Cao2023, Tan2025}. Between these two limits, the logarithmic description passes from routinely valid to absent.

The range of relative roughness between these two limits has barely been touched. Gravel-scale roughness, $k_N = O(1 \mathrm{cm})$ and hence $A_{bm}/k_N \sim O(10$--$100)$, describes gravel beaches, rubble slopes and coarse lag deposits, yet it has received little systematic attention. A few studies in this regime \cite{JonssonCarlsen1976, Dixen2008, Hay2012b, ODonoghue2021} do report near-bed logarithmic profiles, consistent with the sand-scale picture. But here the scale hierarchy is marginal in its lower link — $\delta_w$ is only a few times $k_N$ — so the observed logarithmic layer is squeezed directly against the roughness, and how much of the boundary layer can be affected by roughness elements has never been examined. The densely packed ceramic-marble bed of \citet{Yuan2014, Yuan2015} (YM14 and YM15 hereafter), with marble diameter $D = 12.5\ \mathrm{mm}$, $k_N = 20\ \mathrm{mm}$ and $A_{bm}/k_N \sim O(10$--$100)$, is a well-controlled laboratory realisation of this regime, in effect an idealised gravel bed.

For steady flows, the question of how far the roughness influence extends above the bed has a classical answer. Building on the pipe-flow experiments of \citet{Nikuradse1933} and the systematic geometry variations of \citet{Schlichting1936}, \citet{Raupach1991} established that the direct dynamical influence of the roughness is confined to a roughness-controlled layer (RCL), the near-wall region in which the flow depends explicitly on the length scales of the individual elements, wall similarity does not apply, and the logarithmic law does not strictly hold. Its thickness is set not by the far field of individual element wakes but by a collective instability: the distributed drag within a canopy inflects the mean profile at its top, and the resulting mixing-layer eddies penetrate two to five element heights above the crests\cite{raupach1996coherent, finnigan2000} — a mechanism that presupposes a permeable canopy interior, which a single layer of spheres resting on a wall lacks (\S~\ref{sec:RCL in steady flow}). Laboratory measurements over three-dimensional roughness\cite{oloughlin1965, oloughlin1969, mulhearn1978, Raupach1981, raupach1986} consistently place the RCL top, $k_{\mathrm{RCL}}$, at two to five element heights $k$ above the bed, and field studies broadly reproduce this picture, from $k_{\mathrm{RCL}} \approx 5k$ over sparse, scrub-like vegetation \cite{garratt1980} down to a top sitting just above the crests for tightly packed canopies such as wheat\cite{thom1971}. \citet{Raupach1981} linked the spread to the lateral element dimension, \citet{garratt1980} to the mean inter-element spacing $s$, and \citet{Jimenez2004} synthesised it into a single spacing-dependent estimate. This entire body of evidence, however, is steady: whether it carries over to an oscillatory boundary layer, whose logarithmic region is itself only centimetres thick, has never been examined.

Over the marble bed, the two descriptions contradict each other: YM14 and YM15 fitted logarithmic profiles to within a few millimetres of the crests, yet the steady-flow range $2$--$5k$ puts the top of the RCL $2.5$--$6.3$ cm above the bed, comparable to the entire wave boundary layer. Reconciliation therefore requires an unusually thin RCL - thinner than mean-velocity fitting, blind to the turbulence and the element-scale spatial inhomogeneity, can by itself establish. The present work therefore re-analyses the raw PIV records of YM14 and YM15 with a triple decomposition that separates the coherent dispersive motion of the marble wakes from the stochastic turbulence. \S~\ref{sec:background} reviews the theoretical background; \S~\ref{sec:reanalysis} describes the experiments and the data reduction; \S~\ref{sec:RCL} and \S~\ref{sec:dynamics} quantify the thickness of the layer, the turbulence signatures that distinguish it from the logarithmic region above, and the partition of momentum transfer between dispersive and Reynolds stresses; \S~\ref{sec:discussion} discusses the implications; \S~\ref{sec:conclusions} summarises the conclusions.

\section{Theoretical background}\label{sec:background}
\subsection{The roughness-controlled layer in steady flow}\label{sec:RCL in steady flow}

As summarised in \S~\ref{sec:intro}, steady-flow theory confines the direct dynamical influence of the roughness to an RCL of two to five element heights: the near-wall region in which the flow depends explicitly on the length scales of the individual elements, so that wall similarity does not apply and the logarithmic law does not strictly hold. Operationally, \citet{Raupach1991} marks its top, $k_{\mathrm{RCL}}$, as the elevation where the mean shear first recovers its inertial-sublayer value, the logarithmic sublayer occupying the transition region above. The geometric estimate of \citet{Jimenez2004},
\begin{equation}
\frac{k_{\mathrm{RCL}}}{k} = \min\left(1 + \frac{s}{k},\; 5\right),
\label{eq:jimenez}
\end{equation}
brackets the observations from sparse arrays, for which \citet{sabot1977} found $k_{\mathrm{RCL}}/k \approx 4.5$ for spanwise fences at $s/k = 10$, down to tightly packed geometries. For the densely packed marbles considered here, the centre-to-centre spacing $s \approx k$ places the estimated top near $2k$. This estimate sets the quantitative expectation against which the measured RCL top — denoted $z_{\mathrm{RCL}}$ throughout, with $k_{\mathrm{RCL}}$ reserved for the steady-flow prediction — is tested in \S~\ref{sec:RCL}.

\begin{figure}
  \centering  \includegraphics[width=\columnwidth]{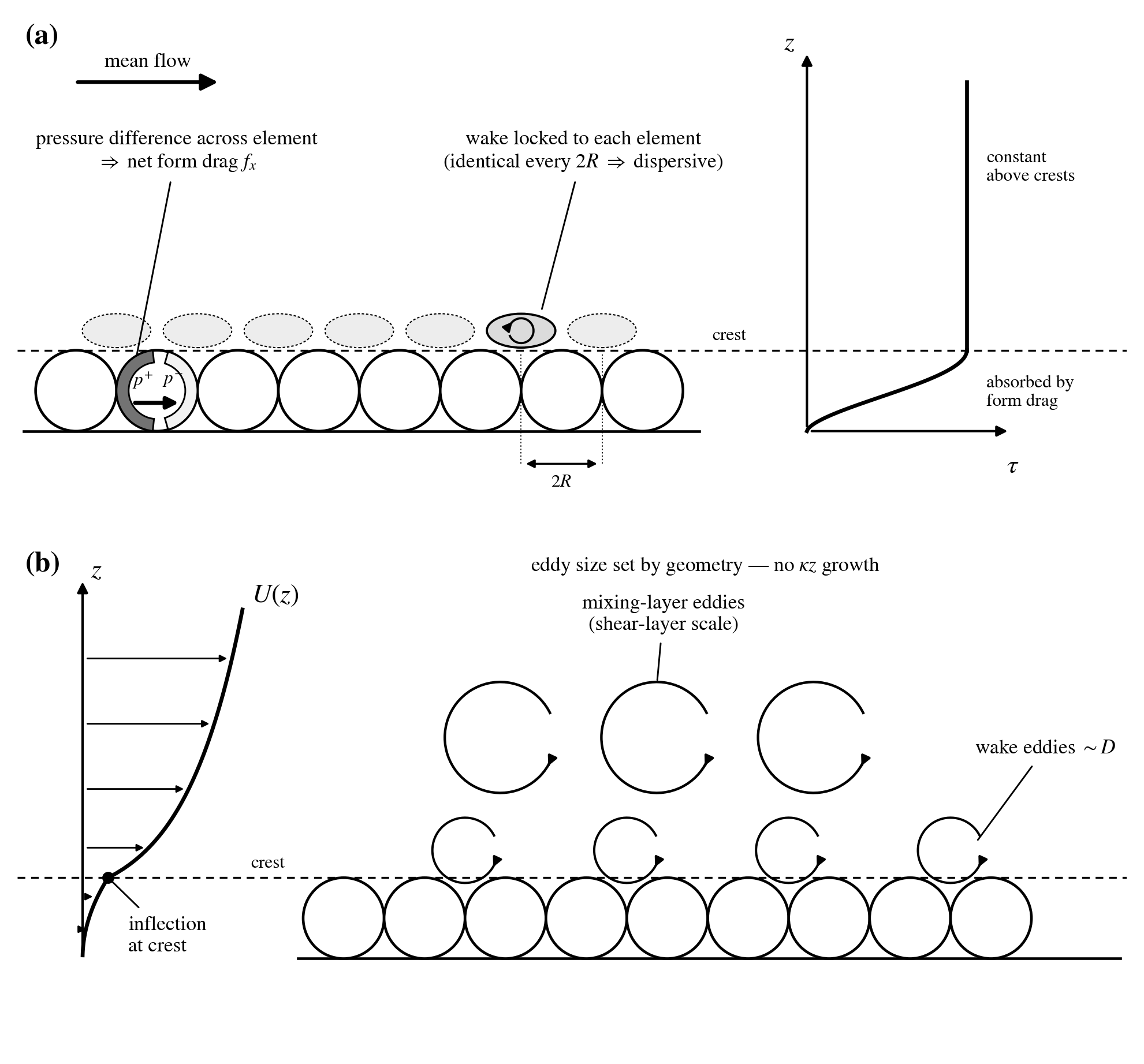}
  \caption{The two mechanisms that break the logarithmic law inside the roughness-controlled layer. (a)~Form drag and the element-locked wake, with the resulting shear stress at right: $\tau$ is constant above the crests and absorbed over the element height below them. (b)~The inflected mean profile at the crest and the two eddy scales it sets.}
  \label{Fig sche}
\end{figure}

Within the RCL, the logarithmic law breaks down on both of its premises. The first concerns the shear stress. In the logarithmic layer, the vertical flux of mean momentum is constant with height and carried entirely by the turbulent Reynolds stress. Inside the RCL, it is not: every element extracts momentum through the pressure difference between its upstream and lee faces (form drag), so the Reynolds stress is progressively reduced towards the bed, while the wake locked behind each crest carries a growing share of the momentum flux as a spatially coherent, dispersive stress\cite{RaupachShaw1982, Raupach1991} (Fig.~\ref{Fig sche} panel a). The shear stress inside the layer therefore has two carriers — Reynolds plus dispersive — with the dispersive share increasing with depth into the layer; the formal definition of the dispersive stress is deferred to \S~\ref{sec:reanalysis}. The second concerns the eddy size. In the logarithmic layer, the distance from the wall, $z$, is the only available length scale, so the energy-containing eddies grow as $\kappa z$. Inside the RCL, the eddies scale with the roughness instead. The kinetic energy the mean flow loses to form drag reappears as wake turbulence at the element scale, while the mean profile, inflected near the crests, feeds mixing-layer (Kelvin-Helmholtz) eddies sized by the inflectional shear layer \cite{raupach1996coherent, finnigan2000, Jimenez2004} (Fig.~\ref{Fig sche} panel b). Either way, the eddy size within the layer is fixed by the roughness geometry, not by $\kappa z$.

Both mechanisms, however, take time to act: the wake must be shed, and its shear layer must roll up, over an eddy turnover time $\tau_e \sim k/u_*$, with $k$ the element height and $u_*$ the shear velocity. In an oscillatory flow, whose forcing reverses every half-period, the steady-flow picture can therefore only apply if the wakes mature well within the period. What the steady-flow picture leaves open is the thickness. The estimate \eqref{eq:jimenez} rests on canopy and sparse-array geometries; a single layer of spheres resting on a wall has no canopy interior, and its spacing is the smallest the geometry permits. What that packing implies for the layer thickness is the subject of \S~\ref{sec:kinematic decay}.

\subsection{Kinematic decay of the dispersive field above a periodic bed}\label{sec:kinematic decay}

Consider first a flow with no wakes at all: attached, irrotational flow over the crests of a bed whose elements repeat with streamwise spacing $s$. Any deviation of the velocity from its bed average is periodic in $x$ with the same spacing, so the deviation potential $\phi$ satisfies the Laplace equation and may be written as
\begin{equation}
\phi(x,z) = \sum_{n\geqslant 1} f_n(z)\cos\!\left(k_n x + \alpha_n\right),
\qquad k_n = \frac{2\pi n}{s},
\label{eq:fourier expansion}
\end{equation}
in which the $n=0$ term is absent because the bed average has been removed. Each amplitude obeys $f_n'' - k_n^2 f_n = 0$, and boundedness far from the bed leaves
\begin{equation}
f_n(z) = C_n\,\mathrm{e}^{-2\pi n z/s} .
\label{eq:mode decay}
\end{equation}

Every harmonic decays exponentially; the higher ones faster, so a fraction of a spacing above the bed, only the $n=1$ mode survives, and the velocity deviation from the bed average dies away over the decay length
\begin{equation}
\ell = \frac{s}{2\pi} .
\label{eq:decay length}
\end{equation}
The strength of the deviation at the crest depends on the element geometry; how quickly it vanishes with height depends on the spacing alone. 

A layer of element-scale spatial variation therefore exists above the crests even in the complete absence of wake dynamics — an apparent RCL that any velocity-based criterion will detect. How thick is it? At crest level, the deviation has some amplitude; write its ratio to the mean flow speed at that height as $R_c$. If the top of the layer is defined as the height at which the ratio first falls below a chosen threshold $R_{\mathrm{thr}}$, then setting $R_c\,\mathrm{e}^{-\Delta z/\ell} = R_{\mathrm{thr}}$ gives
\begin{equation}
\Delta z_{\mathrm{thr}} = \ell \ln\frac{R_c}{R_{\mathrm{thr}}}. \label{eq:threshold height}
\end{equation}
For the close-packed marbles, the spacing equals the diameter, $s \approx D$, so $\ell \approx 0.16D$; with $R_c = 1$ and $R_{\mathrm{thr}} = 0.5$, \eqref{eq:threshold height} predicts a top only $\approx 0.1D$ above the crests: the apparent RCL is only a fraction of $k$.

This apparent RCL, however, carries no dispersive stress. For potential flow over a fore--aft symmetric element, $u''$ is even in $x$ about each crest and $w''$ is odd, so
\begin{equation}
\langle u''w''\rangle = \frac{1}{s}\int_0^{s} u''\,w''\,\mathrm{d}x = 0
\label{eq:zero dispersive stress}
\end{equation}
identically: any substantial $\tau_D$ requires the asymmetry that separation supplies. Two distinct layers are therefore in play. The apparent RCL, defined by the magnitude of the deviation through the dispersive velocity or energy, exists even without wakes. The dynamical RCL, defined by a substantial dispersive stress, is what only separation can produce. \S~\ref{sec:RCL} and \S~\ref{sec:dynamics} measure both, respectively.

\subsection{The near-bed logarithmic region in wave-current flow}\label{sec:BL over a roughwall}

Consider a collinear wave-current bottom boundary layer, with total streamwise velocity $u(z,t)$ driven by a free stream $u_\infty(t) =  U_{\mathrm{orb}}\cos(\omega t)+ U_{\mathrm{curr}}$. In the Grant--Madsen description \cite{Grant1979}, the turbulence is closed with a time-invariant eddy viscosity $\nu_T = \kappa u_{*cw} z$ within the wave boundary layer, where $u_{*cw}$ is the shear velocity based on the maximum combined bed shear stress, and no slip is applied at $z_0 = k_N/30$. The oscillatory component then obeys the deficit equation of \citet{Kajiura1968} and \citet{Grant1979, grant1986continental}, and close to the bed ($z \lesssim l/4$, with $l = \kappa u_{*cw}/\omega$), its first harmonic reduces to a logarithmic profile oscillating in unison,
\begin{equation}
u_w(z,t) \simeq \frac{u_{*wm}^2}{\kappa u_{*cw}}\,\ln\frac{z}{z_0}\,\cos\big(\omega t + \varphi_\tau\big),\label{eq:uw}
\end{equation}
where the bed shear stress leads the free-stream velocity by $\varphi_\tau \approx 10$--$30^\circ$ in the rough turbulent regime. This logarithmic first harmonic is the template for the log-profile fitting of \S~\ref{sec:Region identification}.

The mean current is logarithmic as well. Below the wave-current boundary-layer thickness $\delta_{cw}$, the wave-generated eddies mix the mean momentum and the slope is reduced,
\begin{equation}
\bar{u}_c(z) = \frac{u_{*c}^2}{\kappa u_{*cw}}\,\ln\frac{z}{z_0}, \qquad z < \delta_{cw},\label{eq:Uc mean}
\end{equation}
while above $\delta_{cw}$, the ordinary slope $u_{*c}/\kappa$ is recovered. Under the present conditions, $\delta_{cw}$ exceeds the measurement domain, so the measured current profiles follow the wave-mixed branch \eqref{eq:Uc mean} at every height. Because both components are logarithmic near the bed and anchored at the same $z_0$, they superpose into a single logarithmic profile of time-varying amplitude,
\begin{equation}
u(z,t) \simeq \frac{\ln (z/z_0)}{\kappa u_{*cw}}\,\Big[\, u_{*c}^2 + u_{*wm}^2 \cos\big(\omega t + \varphi_\tau\big) \Big],
\label{eqn:superposed WC}
\end{equation}
whose amplitude is largest when the wave velocity is aligned with the current (the reinforced half-cycle) and smallest when it opposes it. Away from the extrema, this quasi-steady picture degrades, and instantaneous logarithmic behaviour is lost during roughly $20$--$30\%$ of the period around flow reversal\cite{Yuan2014}.

In short, a logarithmic profile exists near the bed of a wave-current boundary layer in three senses: in the instantaneous velocity around the times of maximum velocity, when the flow is quasi-steady; in the first-harmonic amplitude \eqref{eq:uw}; and in the time-averaged velocity \eqref{eq:Uc mean}. 

\section{Re-analysis of oscillating water tunnels (OWT) experiment over marble bed}\label{sec:reanalysis}

\subsection{Brief introduction of OWT experiments by YM14, YM15}\label{sec:facility}

\begin{figure}
 \centering
  \includegraphics[width=\columnwidth]{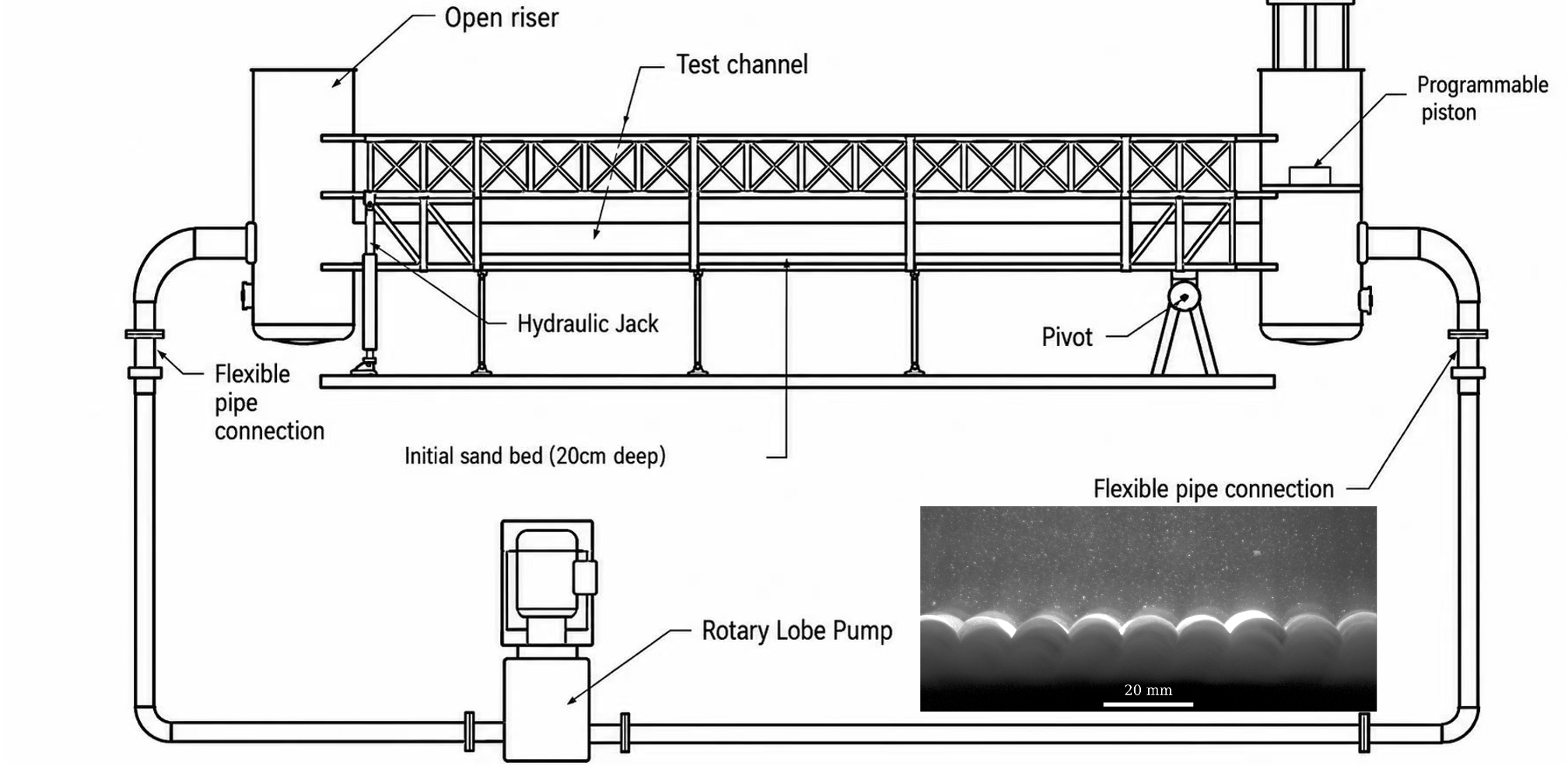}
  \caption{Sketch of the Wave-Current-Sediment (WCS) oscillating water tunnel of YM14 and YM15. The inset shows part of a PIV image near the ceramic-marble bottom.}
\label{Fig Experimental setup}
\end{figure}

The experimental datasets analysed here were obtained by YM14 and YM15 in a full-scale oscillating water tunnel (Fig.~\ref{Fig Experimental setup} panel a). A hydraulically actuated piston drives flow oscillations up to roughly $2$ m excursion amplitude and velocities near $2~\mathrm{m/s}$ in the $10$ m long, $40$ cm $\times$ $50$ cm test channel, whose glass sidewalls and acrylic lids allow optical access for laser-based diagnostics. A separate rotary-lobe pump superimposes a steady current of up to $\sim 60~ \mathrm{cm/s}$, reversible in direction, whose discharge stays constant during wave generation. Further details of the facility are given in YM14 and YM15.

Of the three bed configurations tested (smooth, "sandpaper", ceramic-marble), the present analysis uses the ceramic-marble bottom: a single layer of densely packed, uniformly placed ceramic marbles of diameter $D = 12.5$ mm, giving a well-defined rough surface and a natural vertical reference at the marble crests. By logarithmic-profile fitting of near-bottom velocities from pure current and pure-sinusoidal-wave tests, YM14 placed the theoretical bed origin $z = 0$ at $4.0 \pm 0.4$ mm (about one-third of $D$) below the crests and obtained a Nikuradse equivalent sand-grain roughness $k_N = 20 \pm 3$ mm. Comparable offsets have been obtained independently over similar beds: \citet{Dixen2008} placed the theoretical bed at $0.23D$ below the crests for packed spheres and $0.25D$ for stones, by the same procedure. 

The flow conditions analysed here are drawn from YM15 (table \ref{tab:flow conditions}). Only the sinusoidal wave shape is used, at two periods, $T = 6.25$ and $12.5$ s, resolved respectively into 32 and 64 phase points. The amplitude was fixed through the first-harmonic piston displacement, so the measured $U_{\mathrm{orb}}$ departed from its nominal value by only $1$--$5\%$, and every test was sampled over 32 successive periods to secure reliable phase averaging. A collinear current was superimposed at two pump settings, so that the ratio of current to maximum wave bottom shear stress spanned roughly $8$--$60\%$. The free-stream velocity of a wave-current flow is the summation of the two components,
\begin{equation}
u_\infty(t) = U_{\mathrm{orb}}\cos(\omega t)+ U_{\mathrm{curr}},
\label{eq:wave_shape}
\end{equation}
where $U_{\mathrm{curr}}<0$ in these datasets.
The finding of YM14 and YM15 central to the present study is that a logarithmic profile exists in the very near-bed region, in the first-harmonic amplitude and in the instantaneous velocity around its maxima, and that its lower end reaches within a few millimetres of the marble crests.

\begin{table}
\caption{Target flow conditions. Cases are grouped by flow type: pure wave (W), pure current (C) and wave-current (WC). $a_1$ is the first-harmonic displacement amplitude of the piston, $U_{\mathrm{orb}}$ the first-harmonic (wave) free-stream velocity amplitude, $T$ the wave period, $\overline{U}_{\mathrm{curr}}$ the average current velocity nominal magnitude.}
\label{tab:flow conditions}
\begin{ruledtabular}
\begin{tabular*}{\columnwidth}{@{\extracolsep{\fill}}cl*{4}{d}@{}}
Case number & Flow type    & \mbox{$a_1$} & \mbox{$U_{\mathrm{orb}}$} & \mbox{$T$}  & \mbox{$\overline{U}_{\mathrm{curr}}$} \\
            &              & \mbox{(mm)} & \mbox{(cm/s)}            & \mbox{(s)} & \mbox{(cm/s)} \\
\hline
W1  & pure wave    & 250       & 98.7      & 6.25      & \mbox{--} \\
W2  & pure wave    & 200       & 39.5      & 12.5      & \mbox{--} \\
W3  & pure wave    & 400       & 79.0      & 12.5      & \mbox{--} \\
W4  & pure wave    & 400       & 157.9     & 6.25      & \mbox{--} \\[5pt]
C1  & pure current & \mbox{--} & \mbox{--} & \mbox{--} & 14.82 \\
C2  & pure current & \mbox{--} & \mbox{--} & \mbox{--} & 45.60 \\[5pt]
WC1 & wave-current & 250       & 98.7      & 6.25      & 14.82 \\
WC2 & wave-current & 400       & 157.9     & 6.25      & 14.82 \\
WC3 & wave-current & 250       & 98.7      & 6.25      & 45.60 \\
WC4 & wave-current & 400       & 79.0      & 12.5      & 14.82 \\
WC5 & wave-current & 400       & 157.9     & 6.25      & 45.60 \\
\end{tabular*}
\end{ruledtabular}
\end{table}

\subsection{Data analysis}\label{sec:measurement}

\subsubsection{PIV analysis}\label{sec:PIV analysis}

\begin{figure}
  \centering
  \includegraphics[width=\columnwidth]{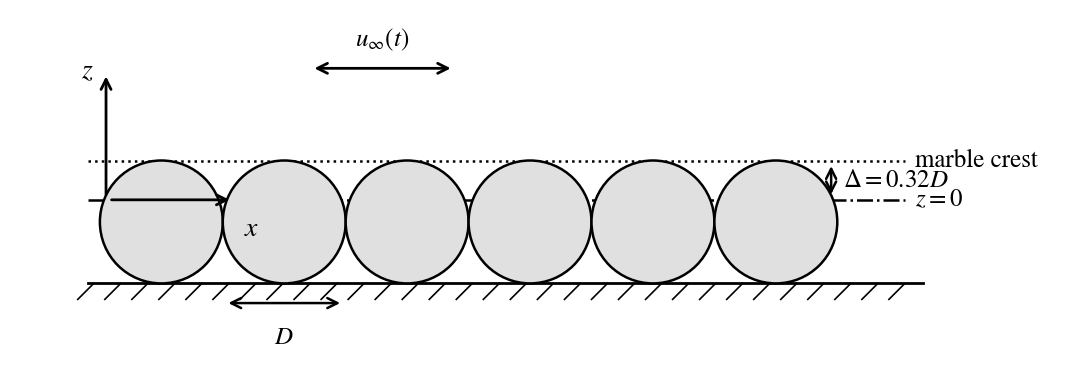}
  \caption{Definition sketch of the vertical coordinate: the theoretical bed origin $z=0$ lies $\Delta=0.32D$ below the marble crests (dotted line).}
  \label{Fig sketch of marble beds}
\end{figure}

We compute the velocity fields directly from the raw image pairs of YM14 and YM15 for the eleven flow conditions of table~\ref{tab:flow conditions}, using an in-house Python algorithm. Each image pair is interrogated on a single-pass grid of $128 \times 8$-pixel windows overlapped $50\%$, finer in the vertical than the $128 \times 16$ grid of the original study (vertical vector spacing $0.20~\mathrm{mm}$, against $0.40~\mathrm{mm}$). Displacements are obtained from the FFT-based cross-correlation peak with three-point Gaussian sub-pixel refinement, and spurious vectors, a few percent, are rejected by normalised median filtering \cite{Westerweel2005} and replaced by the local mean. The resolution 
is modest but sufficient to resolve the main features of the RCL. The vertical spacing of $0.20~\mathrm{mm}$ ($0.016D$) places about ten
measurement points across a layer of thickness $0.1$--$0.2D$. The streamwise spacing of about $3.2~\mathrm{mm}$ ($0.26D$) gives roughly
four points per marble spacing, enough to sample the marble-scale variability that dominates the dispersive field. Sub-marble details and the smallest turbulent eddies are filtered by the finite interrogation window. The zero-plane offset $\Delta$ and roughness $k_N$ determined by YM14 are adopted, as sketched in Fig.~\ref{Fig sketch of marble beds}. The theoretical bed origin $z = 0$ lies $\Delta = 4.0~\mathrm{mm}$ ($0.32D$) below the marble crests, and all elevations are measured from this origin.

The essential difference from the original reduction of YM14 and YM15 is that the full two-dimensional field $\mathbf{u}(x,z,t)$ is retained. It is phase-averaged at fixed $x$ before any spatial averaging, so that the spatial deviation and the turbulent fluctuation remain distinct fields. In their reduction, the fluctuation was defined as the departure from the double-averaged profile, so the resulting Reynolds stress is an equivalent stress that absorbs the form-induced contribution of individual marbles. The averaging operators and the resulting velocity decomposition are defined in \S~\ref{sec:Triple decomposition}.

\subsubsection{Velocity decomposition and averaging operators}\label{sec:Triple decomposition} 


Separating the dispersive and turbulent contributions to the flow requires two averaging operators, defined here together with the velocity decomposition they induce. The instantaneous velocity in the measurement plane is denoted by $\mathbf{u}(x,z,t)=(u,w)$, and the wave phase is $\theta=\omega t$.

For any scalar quantity $\epsilon(x,z,t)$, the \emph{phase average} ($[\cdot]$) at a fixed phase within the wave period $T$ is defined as

\begin{equation}
[\epsilon(x,z,t)] = \frac{1}{N}\sum_{n=0}^{N-1}\epsilon(x,z,t+nT),
\end{equation}
where $N$ is the number of sampled wave cycles, which in our case is 32.

The \emph{bed average} (spatial average, $\langle\cdot\rangle$) is defined along a horizontal line at a given elevation $z$, over all the streamwise grid points $x_1, x_2, \ldots, x_{N_x}$ of the PIV window, which covers an integer number of marble spacings. The bed average is applied to the phase average via

\begin{equation}
\langle[\epsilon]\rangle = \frac{1}{N_x}\sum_{i=1}^{N_x}[\epsilon](x_i,z,\theta).
\end{equation}

Using these operators, the velocity vector $\mathbf{u}$ is decomposed into a phase-averaged part and a turbulent fluctuation,

\begin{equation}
\mathbf{u} = [\mathbf{u}] + \mathbf{u}',
\end{equation}
and the phase-averaged component is further decomposed into a bed-averaged component and a spatial deviation (dispersive component),

\begin{equation}
[\mathbf{u}] = \langle[\mathbf{u}]\rangle + \mathbf{u}'',
\end{equation}
so that the full decomposition is

\begin{equation}
\mathbf{u} = \langle[\mathbf{u}]\rangle + \mathbf{u}'' + \mathbf{u}'.
\end{equation}

The \emph{time average} (overbar) is taken over one wave period and is applied to phase-averaged or bed-and-phase-averaged quantities. Because every harmonic integrates to zero over a period, it removes the oscillatory part and leaves the steady current: the mean velocity profile is $u_0=\overline{\langle[u]\rangle}$.

Finally, the phase-and-bed-averaged velocity is Fourier analysed over the wave cycle,

\begin{equation}
\langle[u]\rangle(z,\theta) = u_0(z) + \sum_{m\geqslant 1} U_m(z)\cos\big(m\theta - \varphi_m(z)\big),
\end{equation}
where $U_1(z)$ is the first-harmonic velocity amplitude, used for the log-law fitting in \S~\ref{sec:Region identification}. The first-harmonic amplitude of any other phase-averaged, bed-averaged quantity (e.g., the shear stresses of \S~\ref{sec:turbulence stats}) is defined in the same way.

\subsubsection{Turbulence statistics and dispersive quantities}\label{sec:turbulence stats}

Planar PIV resolves the streamwise and wall-normal fluctuations $u'$ and $w'$, so the turbulence is characterised by 

\begin{equation}
k_t(x,z,\theta) = \tfrac{1}{2}\big[\,u'^2 + w'^2\,\big],
\end{equation}
a two-component surrogate for the full turbulent kinetic energy. The corresponding measure of the energy held in the spatial deviation is the dispersive kinetic energy,

\begin{equation}
k_d(z,\theta) = \tfrac{1}{2}\big\langle\, u''^2 + w''^2 \,\big\rangle.
\end{equation}

The phase-averaged kinematic shear stresses (per unit fluid density) carried by the two fields are the Reynolds shear stress and the dispersive shear stress,

\begin{equation}
\tau_\mathrm{{Re}}(z,\theta) = -\big\langle\big[\,u'w'\,\big]\big\rangle,
\qquad
\tau_{D}(z,\theta) = -\big\langle\, u''w'' \,\big\rangle,
\label{eq:RS and DS}
\end{equation}
whose first-harmonic amplitudes, $\tau_{\mathrm{Re,1}}$ and $\tau_{D,1}$, are compared in \S~\ref{sec:momentum transfer}.

The size of the energy-containing eddies is measured by two-point spatial correlations. A two-point correlation requires the statistics to be homogeneous along the separation direction. The mean shear breaks that homogeneity in the wall-normal direction, so separations are taken in the streamwise direction only, and the streamwise fluctuation $u'$ is correlated as the component carrying the energy-containing structures. The phase-averaged and bed-averaged autocorrelation is

\begin{equation}
\Gamma_{uu}(\Delta x, z, \theta)
= \frac{\big\langle\big[\,u'(x,z,t)\,u'(x{+}\Delta x,z,t)\,\big]\big\rangle}
       {\big\langle\big[\,u'^2\,\big]\big\rangle},\label{eq:Ruu}
\end{equation}
where the normalisation uses the approximate streamwise homogeneity of the turbulent variance $\langle u'^2\rangle$ over the bed. This homogeneity is of the turbulent fluctuation after the coherent dispersive field has been removed, and does not contradict the streamwise inhomogeneity of the phase-averaged flow that the dispersive decomposition captures. It is the intensity of the residual turbulence, not the mean or coherent motion, that is taken uniform in $x$. Similar averaging is used by \citet{Tomkins2003}, \citet{Ganapathisubramani2005} and \citet{Volino2007}. The streamwise integral length scale is obtained by integrating the autocorrelation from zero separation to its first zero crossing,

\begin{equation}
L_x(\theta, z) = \int_0^{\Delta x_0} \Gamma_{uu}(\Delta x, z, \theta)~\mathrm{d}(\Delta x),\label{eq:Lx definition}
\end{equation}
where $\Delta x_0(z,\theta)$ is the smallest separation at which $\Gamma_{uu}=0$, or the maximum resolved separation if no zero is reached, in which case $L_x$ is a lower bound.

\subsubsection{Region identification} \label{sec:Region identification}

Three elevations are detected from the reduced fields: the tops of the apparent and dynamical roughness-controlled layers, and the boundaries of the logarithmic profile layer.

The top of the RCL is detected from the ratio of the dispersive velocity magnitude to the mean flow speed,

\begin{equation}
R(z,\theta)
= \frac{\big\langle\, u''^2 + w''^2 \,\big\rangle^{1/2}}
       {\overline{\big\langle\big[\,(u^2+w^2)^{1/2}\,\big]\big\rangle}},
       \label{eq:RCL top threshold}
\end{equation}
in which the numerator is the bed-RMS dispersive velocity magnitude at phase $\theta$ and the denominator is the time- and bed-averaged flow speed. This is the ratio whose crest-level value was written $R_c$ in \S~\ref{sec:kinematic decay}. Physically, $R(z,\theta)$ measures how strongly affected the flow at elevation $z$ is by the individual marbles. Where $R$ is of order one, the wake-induced spatial variability is comparable to the mean motion and the flow is directly organised by the roughness elements. Where $R$ is small, the marble signature has decayed, and the flow has become streamwise homogeneous. $R(z,\theta)$ therefore defines a phase-dependent top of the apparent RCL, $z_{\mathrm{RCL}}(\theta)$, the first height above the crests at which $R(z,\theta) < R_{\mathrm{thr}}$. The period average ratio $\overline{R}(z)$ then gives a single representative top $z_{\mathrm{RCL}}^{R}$, the first height at which $\overline{R}(z)<R_{\mathrm{thr}}$. For definiteness, we take $R_{\mathrm{thr}}=0.5$, and the sensitivity of the top to this choice is quantified in table~\ref{tab:RCL top}. Because the marble-periodic part of the dispersive field decays no more slowly than the $n=1$ mode (\S\,\ref{sec:kinematic decay}), a top detected this way is bounded above by \eqref{eq:threshold height} insofar as the measured deviation is marble-periodic. Contributions at larger scales are not so bounded, and the comparison is made in \S\,\ref{sec:RCL}.

The dispersive stress provides a second, independent detection of the dynamical layer rather than the apparent one. Forming the ratio of the dispersive to the Reynolds stress from the first-harmonic amplitudes of \eqref{eq:RS and DS} (from the time-averaged stresses for the pure current cases), we take the top of the dynamical layer, $z_{0.5}$, as the first height above the crests at which the ratio falls below $0.5$ and remains below it. The threshold parallels $R_{\mathrm{thr}}$ and is likewise a convention. Unlike $R$, however, the quantity it is applied to receives no contribution from a symmetric attached flow of \eqref{eq:zero dispersive stress}, so $z_{0.5}$ shows only what separation produces. The two tops are compared case by case in \S~\ref{sec:momentum transfer}.

Above these two near-crest markers, the boundaries of the logarithmic profile layer (LPL) are identified from the first-harmonic amplitude $U_1(z)$. To avoid defining the window through the shear velocity it is used to measure, the window is set in two stages. A preliminary log-law fit to $U_1(z)$ over the near-bottom data below the first-harmonic overshoot yields an initial shear velocity, and hence an initial length scale $l_w=\kappa u_{*w}/\omega$. $U_1(z)$ is then refitted over $z\in[11~\mathrm{mm},\,0.25\,l_w]$, to give the wave shear velocity $u_{*w}$ and the reference slope $u_{*w}/\kappa$. The local slope is evaluated by central differences on the measurement heights $z_i$ and smoothed by a centred five-point moving average,

\begin{equation}
S(z_i) = \frac{1}{5}\sum_{k=-2}^{2}\left(\frac{\mathrm{d}U_1}{\mathrm{d}\ln z}\right)_{i+k},
\end{equation}
and its relative deviation from the reference slope is

\begin{equation}
\mathrm{dev}(z_i) = \frac{\left|S(z_i)-u_{*w}/\kappa\right|}{u_{*w}/\kappa}.
\end{equation}

The lower boundary $z_{\mathrm{log,b}}$ is the lowest elevation at which $\mathrm{dev}(z)<\mathrm{tol}$ for five consecutive points, and the upper boundary $z_{\mathrm{log,t}}$ is the first elevation above $z_{\mathrm{log,b}}$ at which $\mathrm{dev}(z)>\mathrm{tol}$ for five consecutive points, with $\mathrm{tol}=0.30$. For the pure current cases, there is no first harmonic, and the identical slope-deviation construction is applied to the bed and time-averaged velocity $u_0(z)=\overline{\langle u\rangle}$ in place of $U_1(z)$. A steady current has no oscillatory overshoot and no wave-imposed length scale $l$, so the $0.25\,l_w$ window of the oscillatory cases does not apply. Moreover, the reference slope $u_{*c}/\kappa$ is taken directly from a logarithmic fit to $u_0(z)$, with both $u_{*c}$ and $z_0$ free. The five-point slope smoothing, the five-consecutive-point rule, and the tolerance $\mathrm{tol}=0.30$ are unchanged, and yield $z_{\mathrm{log,b}}$ and $z_{\mathrm{log,t}}$ exactly as in the oscillatory cases. These two boundaries respond only weakly to the choices made in the detection procedure. Taking the reference slope from the log fit itself rather than anchoring it on the independently measured roughness of \citet{Yuan2014} ($k_N = 20~\mathrm{mm}$) moves them by $0.02$--$0.05D$, and changing the smoothing width by less than $0.2D$. The tolerance matters more: as $\mathrm{tol}$ varies over $[0.2, 0.4]$, $z_{\mathrm{log,b}}$ spans $0.80$--$1.06D$ and $z_{\mathrm{log,t}}$ spans $0.87$--$1.27D$. The logarithmic-layer extent is therefore reported as approximate. The relative ordering of the RCL and the LPL is robust, but the gap between them can change quantitatively.

\section{Kinematic characteristics of the apparent RCL}\label{sec:RCL}

Under oscillatory forcing, the RCL cannot be a fixed feature of the flow: the wakes that constitute it are destroyed and rebuilt every half-cycle, so the layer can only be located statistically. \S~\ref{sec:RCL top} measures its thickness and phase evolution through the dispersive-to-mean velocity ratio, and \S~\ref{sec:TPC} the streamwise size of the energy-containing turbulent eddies through two-point correlations of the turbulent fluctuation. 

\subsection{The thickness and phase evolution of the apparent RCL}\label{sec:RCL top}

We begin with the thickness of the layer and how it evolves over the wave cycle. At each phase, the top of the layer, $z_{\mathrm{RCL}}(\theta)$, is detected as in \S~\ref{sec:Region identification} — the first height above the crests at which the dispersive-to-mean ratio $R(z,\theta)$ falls below $R_{\mathrm{thr}} = 0.5$ — and the period-averaged ratio gives the representative top $z^R_{\mathrm{RCL}}$. The sensitivity of the results to this choice is examined in table~\ref{tab:RCL top}, after the measurements themselves.

\begin{figure}
 \centering
  \includegraphics[width=\columnwidth]{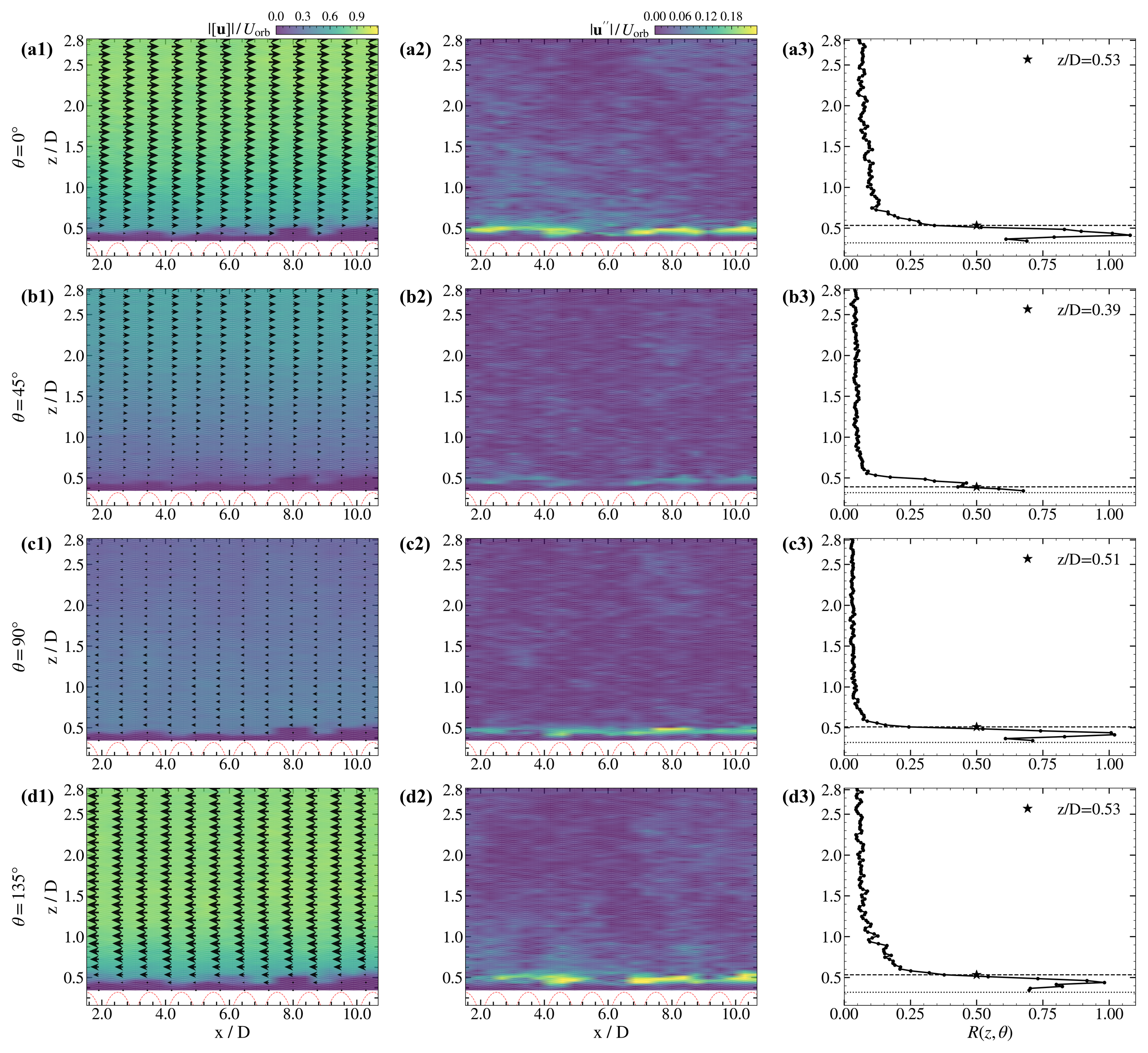}
  \caption{Phase-averaged flow structure of wave-current case WC1 at (a)~$\theta=0^\circ$, (b)~$45^\circ$, (c)~$90^\circ$ and (d)~$135^\circ$. Left: phase-averaged velocity magnitude, with arrows giving in-plane direction only. Centre: dispersive velocity magnitude. Right: the ratio $R(z,\theta)$ of equation~\eqref{eq:RCL top threshold}; the star and dashed line mark $z_{\mathrm{RCL}}$, where $R(z,\theta)$ first falls below $0.5$. Dotted lines mark the marble crests.}
\label{Fig WC contours wave current}
\end{figure}

\begin{figure}
 \centering
  \includegraphics[width=\columnwidth]{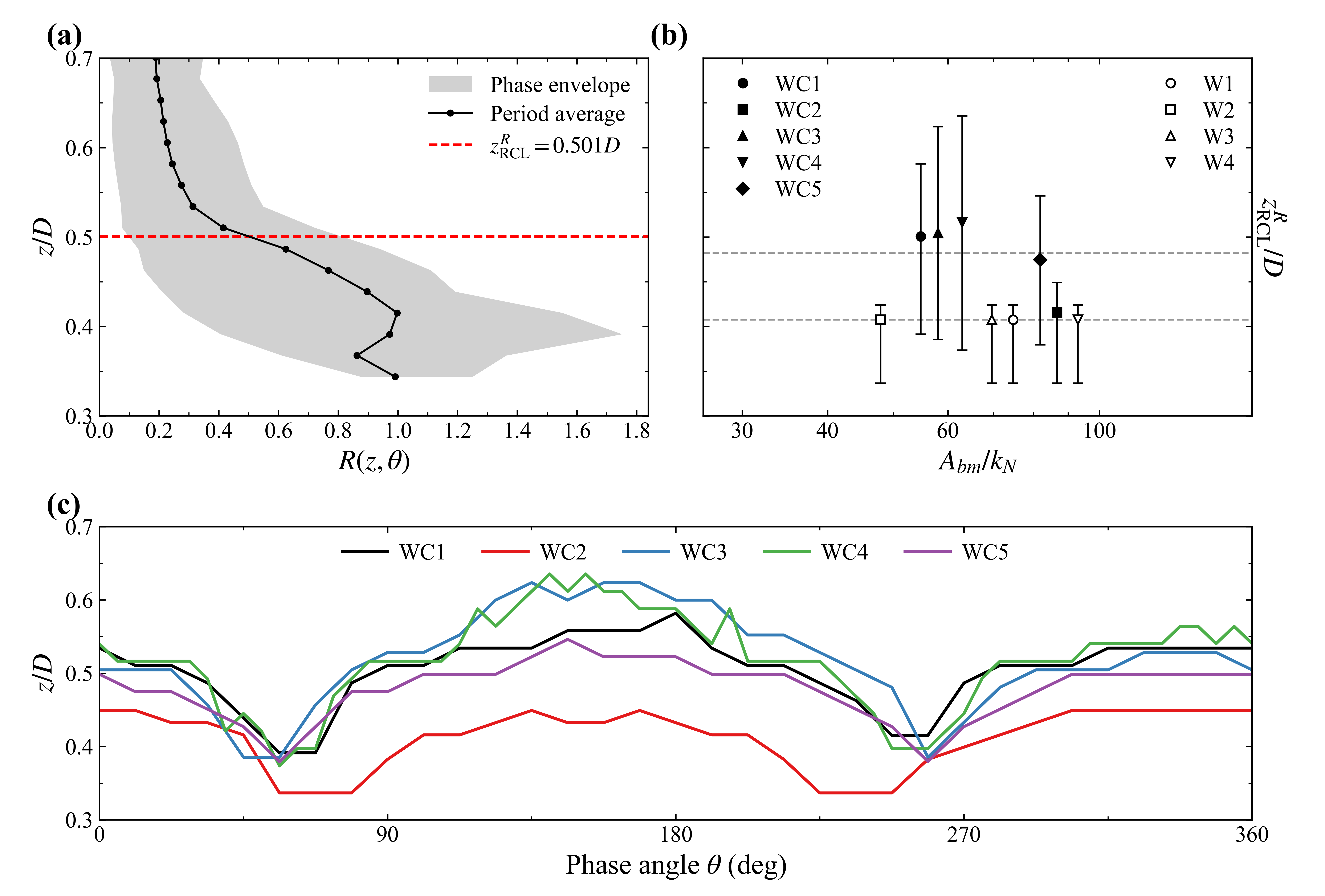}
  \caption{RCL top for the wave-forced cases. (a)~Phase-averaged ratio $R(z,\theta)$ of equation~\eqref{eq:RCL top threshold} for WC1, with its phase envelope, period average and period-averaged top $z^{R}_{\mathrm{RCL}}$. (b)~$z^{R}_{\mathrm{RCL}}/D$ against $A_{bm}/k_N$, wave-current (filled) and pure wave (open); bars span the within-cycle range, grey dashed lines the group averages. (c)~$z_{\mathrm{RCL}}(\theta)/D$ for the five wave-current cases.}
\label{Fig RCL top for different phases}
\end{figure}

\begin{figure}
  \centering
  \includegraphics[height=8cm]{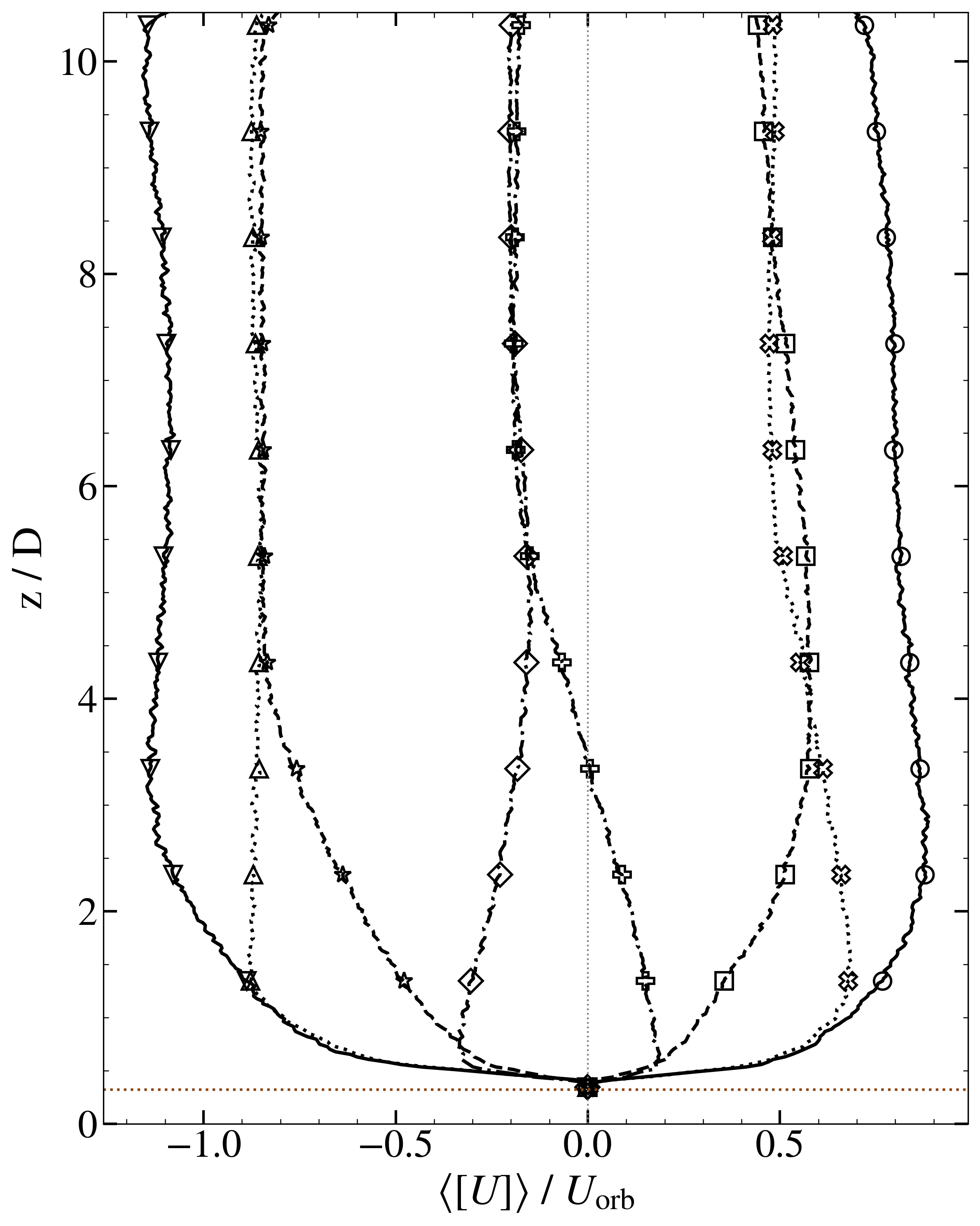}
  \caption{Phase- and bed-averaged streamwise velocity profiles at eight phases of the wave cycle for WC1: $\circ$, $\omega t=0^\circ$; $\square$, $45^\circ$; $\Diamond$, $90^\circ$; $\triangle$, $135^\circ$; $\triangledown$, $180^\circ$; $\ast$, $225^\circ$; $+$, $270^\circ$; $\times$, $315^\circ$. Dotted lines mark the marble crest (horizontal) and zero velocity (vertical).}
  \label{Fig 8 phases free stream}
\end{figure}

Fig.~\ref{Fig WC contours wave current} shows the fields underlying this detection, for the representative wave-current case WC1 at four phases from the free-stream velocity peak through the reversal ($\theta = 0^\circ$, $45^\circ$, $90^\circ$ and $135^\circ$; rows). The left column shows the phase-averaged velocity magnitude, nearly uniform immediately above the crests and reversing direction across $90^\circ$. The centre column shows the dispersive velocity magnitude: a thin band wrapped around the crests (dispersive band hereafter), repeating marble by marble at fixed streamwise positions and therefore tied to the bed geometry, strongest near $z/D \approx 0.5$ and decaying rapidly above. The band is strong at $0^\circ$, all but extinguished at $45^\circ$, and fully re-established by $90^\circ$. The right column combines the two fields into the ratio $R(z,\theta)$, whose crossing of $R_{\mathrm{thr}} = 0.5$ (star) gives the top: $z_{\mathrm{RCL}}(\theta)$ sits near $0.5D$ at $0^\circ$, $90^\circ$ and $135^\circ$ and drops to $0.39D$ at $45^\circ$, the phase at which the dispersive band fades.

To quantify this phase evolution and compare it across flow conditions, Fig.~\ref{Fig RCL top for different phases} presents the detected RCL top for all wave-forced cases. For WC1 (panel a), the envelope of $R(z,\theta)$ over the cycle is widest just above the crests and collapses above $z \approx 0.6D$: whatever the wave does to the near-crest flow happens within the layer itself. Averaged over the period (panel b) for all flow cases, the tops show no systematic trend with $A_{bm}/k_N$: the pure waves cluster at $0.41D$, the wave-current cases sit about $0.1D$ higher near $0.51D$, and WC2, whose current is weakest relative to its wave, falls within the pure wave cluster. Resolved in phase (panel c), the tops of all five wave-current cases repeat one shape: thinnest in two narrow windows near $50$--$70^\circ$ and $245$--$265^\circ$, ahead of the nominal reversals at $90^\circ$ and $270^\circ$, and thickest through the middle of the reinforced half-cycle near $135$--$180^\circ$ (Fig.~\ref{Fig 8 phases free stream}), with WC2 showing the least variation. The lead is consistent with the phase advance of the near-bed velocity over the free stream (\S~\ref{sec:BL over a roughwall}), so the layer thins when the near-bed flow, not the free stream, reverses.

The absolute thickness is the central fact. Across all eleven cases, the top lies only 0.1--0.2$D$ above the crests. This is what the kinematics of \S~\ref{sec:kinematic decay} anticipates: with the crest-level ratio read directly from the measured profiles ($R_c \approx 0.96$--$1.62$ for the eleven cases) and $\ell \approx 0.16D$, \eqref{eq:threshold height} predicts tops of 0.1--0.19$D$, matching the measurements closely — and lying an order of magnitude below the canopy estimate of \eqref{eq:jimenez} for the same geometry.

\begin{figure}
 \centering
  \includegraphics[width=\columnwidth]{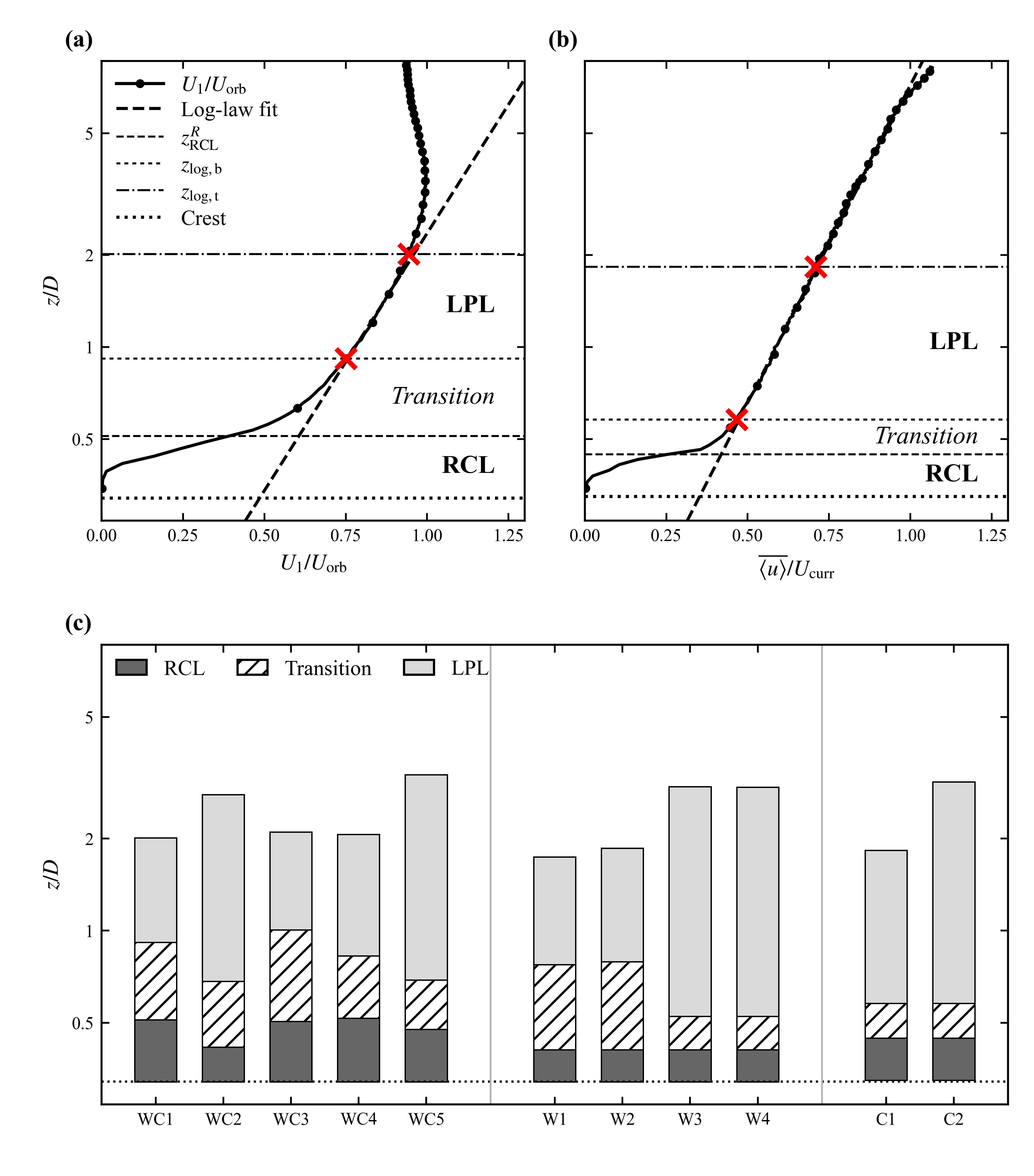}
  \caption{The three-region structure. (a)~First-harmonic velocity amplitude $U_1/U_{\mathrm{orb}}$ for WC1, with the log-law fit (bold dashed) and the region boundaries: RCL top $z^{R}_{\mathrm{RCL}}$ (dashed), LPL bottom $z_{\mathrm{log,b}}$ (fine dotted), LPL top $z_{\mathrm{log,t}}$ (dash-dotted) and the crest (bold dotted); red crosses mark the profile at the LPL boundaries. (b)~The same for the time-averaged velocity $\overline{\langle u\rangle}/U_{\mathrm{curr}}$ of C1. (c)~Extent of the three regions above the crest for all eleven cases, grouped by forcing type; bar tops mark $z_{\mathrm{log,t}}$.}
\label{Fig three region schematic1}
\end{figure}

Fig.~\ref{Fig three region schematic1} places the thin layer within the boundary layer as a whole. For WC1 (panel a), the first-harmonic amplitude follows the fitted logarithm closely between $z_{\mathrm{log,b}}$ and $z_{\mathrm{log,t}}$, bends toward the free-stream overshoot above, and collapses steeply below the RCL top. The interval between the RCL top and $z_{\mathrm{log,b}}$, where the profile has left the logarithm but not yet begun that steep collapse, is the transition region. The time-averaged profile of the pure current C1 (panel b) returns the same arrangement without any wave forcing. Across all eleven cases (panel c), the three-region structure is universal: a roughness-controlled layer of 0.1--0.2$D$, a transition region with no clean group pattern, and a logarithmic profile layer occupying the bulk of the measured range, up to 1.8--2.6$D$.

Table~\ref{tab:RCL top} lists the period-averaged tops obtained with thresholds from 0.1 to 0.6. The pure wave and pure current tops are threshold-independent: every choice returns essentially the same height, because their $\overline{R}(z)$ drops steeply at the layer top. The wave-current $\overline{R}(z)$ instead settles to a level of 0.15--0.35 above the layer and approaches zero only gradually, so the detected top moves with the threshold, and at 0.1 (WC1--WC4) or 0.15 (WC2), no top is found within the measurement domain. Over the range 0.3--0.6, however, the inferred thickness stays at 0.1--0.3$D$ and the ordering of the flow types is preserved. $R_{\mathrm{thr}} = 0.5$ is adopted throughout.

In summary, the apparent RCL is thin in every flow examined: its top lies 0.1--0.2$D$ above the crests, at the scale the packing sets through \eqref{eq:threshold height}, with no trend in $A_{bm}/k_N$ over the range sampled. The clearest systematic departure is produced by the wave-current interaction. The superimposed current not only raises the period-averaged top by about $0.1D$, but also keeps $\overline{R}(z)$ elevated well above the layer: the combined forcing extends the influence of the roughness farther into the flow than either the wave or the current alone.

\begin{table*}
\caption{Period-averaged RCL top normalised by the marble diameter ($z_{\mathrm{RCL}}^{R}/D$, with $D=12.5~\mathrm{mm}$) for all datasets using different threshold ratios $R_{\mathrm{thr}}$. Entries marked '--' indicate that $\overline{R}(z)$ never falls below the threshold within the measurement domain, so no RCL top is defined.}
\label{tab:RCL top}
\footnotesize
\setlength{\tabcolsep}{4pt}
\renewcommand{\arraystretch}{1.15}
\begin{ruledtabular}
\begin{tabular*}{\textwidth}{@{\extracolsep{\fill}}c*{11}{d}@{}}
 & \multicolumn{11}{c}{$R_{\mathrm{thr}}$} \\
\cline{2-12}
Case number & 0.1 & 0.15 & 0.2 & 0.25 & 0.3 & 0.35 & 0.4 & 0.45 & 0.5 & 0.55 & 0.6 \\
\hline
W1  & 0.441     & 0.441 & 0.441 & 0.424 & 0.424 & 0.424 & 0.424 & 0.407 & 0.407 & 0.407 & 0.407 \\
W2  & 0.458     & 0.441 & 0.441 & 0.424 & 0.424 & 0.424 & 0.424 & 0.407 & 0.407 & 0.407 & 0.407 \\
W3  & 0.458     & 0.441 & 0.441 & 0.424 & 0.424 & 0.424 & 0.424 & 0.407 & 0.407 & 0.407 & 0.407 \\
W4  & 0.441     & 0.441 & 0.424 & 0.424 & 0.424 & 0.424 & 0.424 & 0.407 & 0.407 & 0.407 & 0.407 \\
C1  & 0.478     & 0.462 & 0.462 & 0.462 & 0.462 & 0.462 & 0.445 & 0.445 & 0.445 & 0.445 & 0.445 \\
C2  & 0.495     & 0.478 & 0.462 & 0.462 & 0.462 & 0.462 & 0.462 & 0.462 & 0.445 & 0.445 & 0.445 \\
WC1 & \mbox{--} & 0.915 & 0.677 & 0.558 & 0.558 & 0.534 & 0.534 & 0.510 & 0.501 & 0.501 & 0.501 \\
WC2 & \mbox{--} & \mbox{--} & 1.182 & 0.499 & 0.449 & 0.449 & 0.432 & 0.432 & 0.416 & 0.416 & 0.416 \\
WC3 & \mbox{--} & 1.218 & 0.718 & 0.623 & 0.576 & 0.552 & 0.528 & 0.528 & 0.505 & 0.505 & 0.505 \\
WC4 & \mbox{--} & 1.492 & 0.826 & 0.659 & 0.588 & 0.564 & 0.540 & 0.516 & 0.516 & 0.516 & 0.516 \\
WC5 & 1.546     & 0.665 & 0.546 & 0.522 & 0.498 & 0.498 & 0.498 & 0.498 & 0.474 & 0.474 & 0.474 \\
\end{tabular*}
\end{ruledtabular}
\end{table*}

\subsection{Streamwise scales of the near-crest structures}\label{sec:TPC}

\begin{figure}
 \centering
\includegraphics[width=\columnwidth]{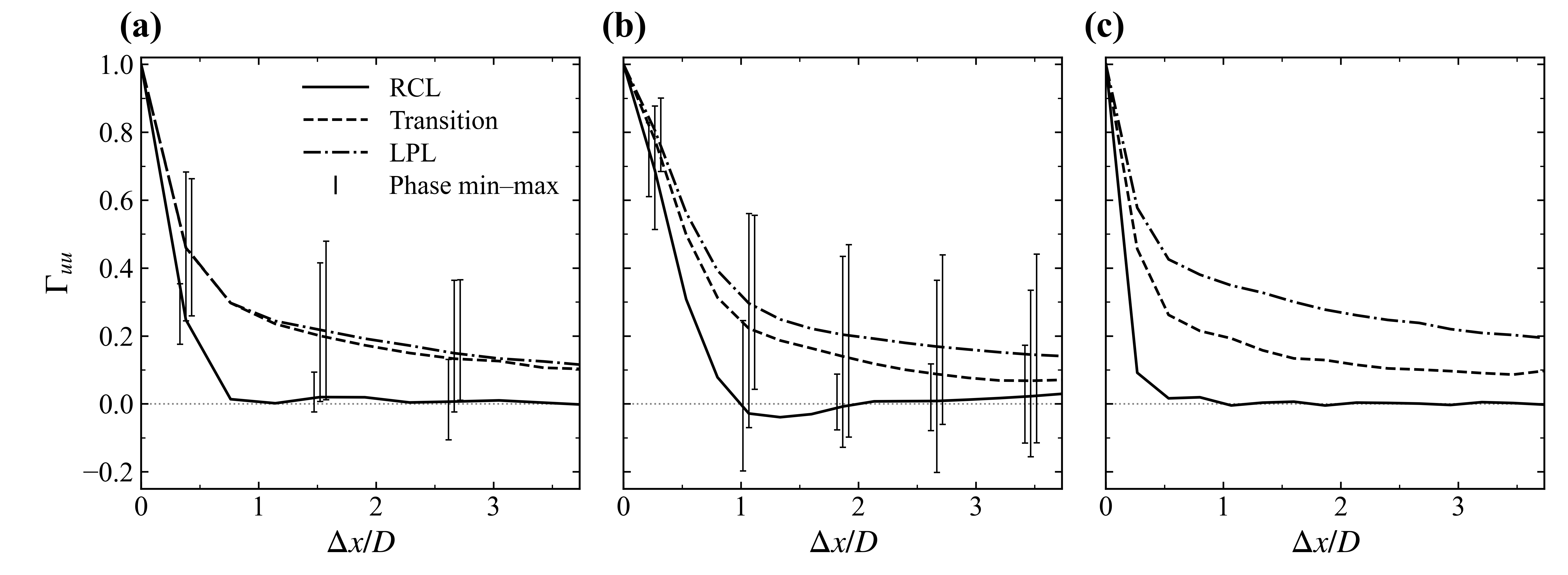}
  \caption{Period-averaged two-point correlations $\Gamma_{uu}(\Delta x)$ in the RCL (solid), transition region (dashed) and LPL (dash-dotted), for (a)~WC3, (b)~W3 and (c)~C2, time-averaged in (c). Bars span the phase range, and the dotted line marks $\Gamma_{uu}=0$.}
\label{Fig Ruu 4 phases}
\end{figure}

\begin{figure}
 \centering
\includegraphics[width=\columnwidth]{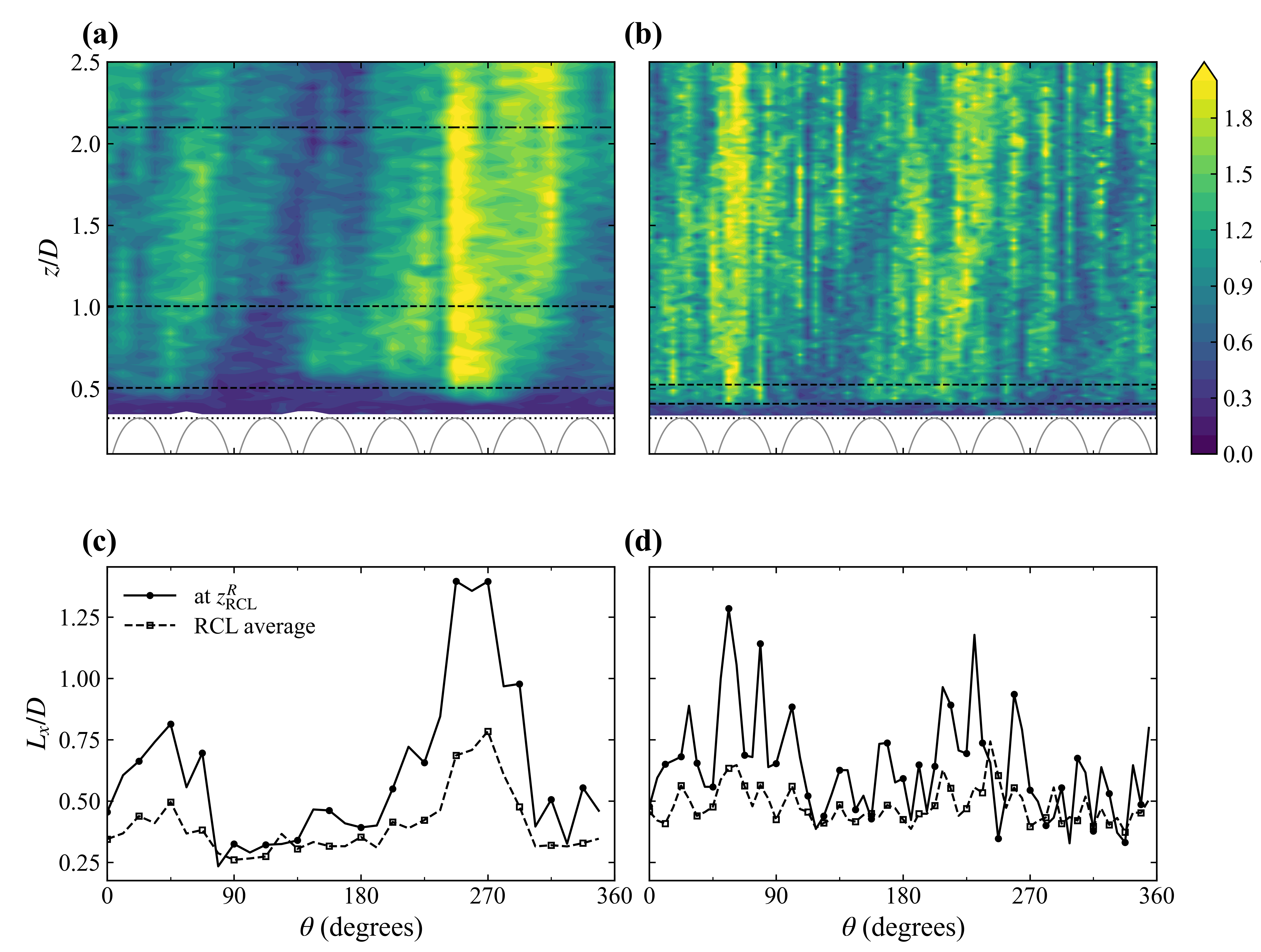}
   \caption{Phase-averaged streamwise integral length scale for (a,c)~WC3 and (b,d)~W3. (a,b)~$L_x(\theta,z)/D$ on a common colour scale, saturated above $L_x/D=2$ (pointed cap). Horizontal lines mark, in ascending order, the crest (dotted), $z^{R}_{\mathrm{RCL}}$, $z_{\mathrm{log,b}}$ and $z_{\mathrm{log,t}}$, with the bed sketched below. (c,d)~$L_x/D$ at the RCL top (solid) and averaged over the RCL (dashed) against phase.}
\label{Fig integral length}
\end{figure}

\begin{figure}
 \centering
\includegraphics[width=\columnwidth]{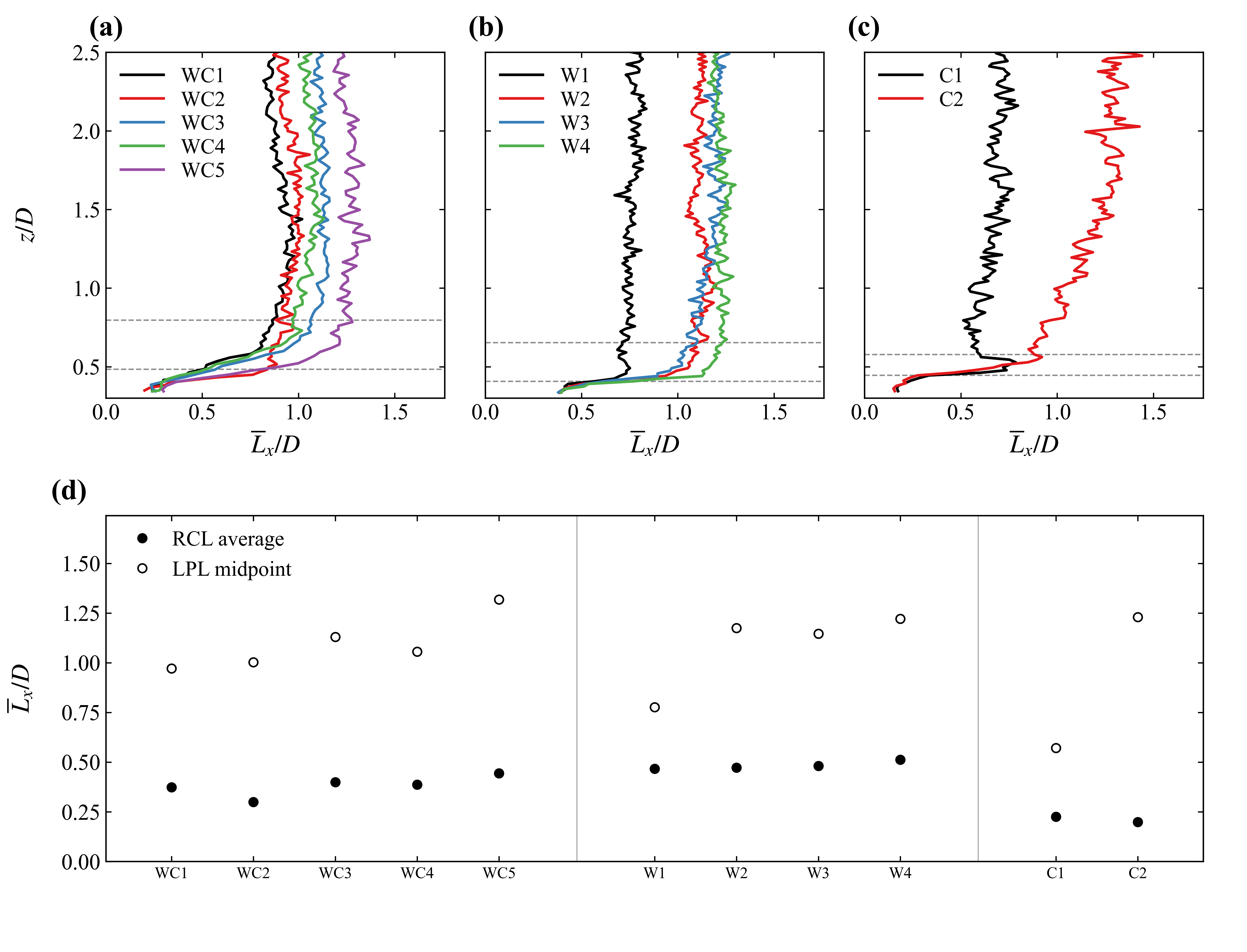}
   \caption{Period-averaged (time-averaged for the pure current cases) streamwise integral length scale for all eleven cases. (a--c)~Profiles $\overline{L}_x(z)/D$ for the wave-current, pure wave and pure current groups. Grey dashed lines mark, in ascending order, the group-mean $z^{R}_{\mathrm{RCL}}$ and $z_{\mathrm{log,b}}$. (d)~Per-case summary. Vertical lines separate the groups.}
\label{Fig integral length summary}
\end{figure}

The turbulence within and above the RCL is characterised by the streamwise size of the energy-containing eddies. At each elevation, the autocorrelation of the streamwise turbulent fluctuation, $\Gamma_{uu}(\Delta x,z,\theta)$ of \eqref{eq:Ruu}, is computed at every phase separately and then averaged over the period (for the pure current cases, over time). Fig.~\ref{Fig Ruu 4 phases} shows the period-averaged (and time-averaged) correlations, sampled at the geometric midpoints of the three regions (RCL, transition region and LPL), for one case of each forcing type: wave-current WC3 (panel a), pure wave W3 (panel b) and pure current C2 (panel c). In every case, the correlation decays with $\Delta x$ most rapidly at the RCL height and progressively more slowly in the transition region and the LPL: the eddies are smallest near the crest and enlarge with height, and the pure current correlation at the RCL height falls off within a fraction of a diameter, the shortest of any case. The vertical bars in panels (a,b), spanning the full range of the phase-averaged correlation, are substantial and at some $\Delta x$ exceed the period-averaged value itself, so the eddy scale varies strongly through the wave cycle. Each phase is therefore reduced to a single number, the integral length $L_x(\theta, z)$ of \eqref{eq:Lx definition}, to characterise the instantaneous eddy size.

Fig.~\ref{Fig integral length} resolves the eddy size in phase, for the wave-current case WC3 (panels a,c) and the pure wave W3 (panels b,d). Panels (a,b) map $L_x(\theta,z)$ over the wave cycle, with the layer boundaries of Fig.~\ref{Fig three region schematic1} marked. The striking feature is a pair of bright vertical bands: phases at which the eddies lengthen at all heights, from the RCL through the transition region into the LPL. Panels (c,d) quantify these bands within the layer, plotting $L_x$ at the RCL top and its average over the RCL against phase. After each reversal, $L_x$ falls back and stays low, near $0.2$--$0.3D$, through the acceleration — the quarter-cycle from the reversal to the following velocity extrema ($\theta = 90$--$180^\circ$ and $270$--$360^\circ$). Through each deceleration — extrema to reversal ($0$--$90^\circ$ and $180$--$270^\circ$) — the eddies lengthen, $L_x$ peaking near $\theta \approx 45^\circ$ and $248^\circ$, one peak in each deceleration window. For the pure wave, the two peaks are almost equal ($\approx 1.28D$), as the symmetry of the forcing requires. When adding the current, they are markedly unequal, $1.40D$ in the reinforced window against $0.85D$ in the opposing one. Set against Fig.~\ref{Fig RCL top for different phases}(c), the timing is sharp: in both windows the $L_x$ maxima precede the thickness minima by $\approx 11^\circ$ — the eddies are longest just before the layer is thinnest.

Fig.~\ref{Fig integral length summary} completes the picture with the period average. Panels (a--c) show the profiles $\overline{L}_x(z)$ for the wave-current, pure wave and pure current groups, with the group-mean layer boundaries marked, and all eleven cases share one shape: below the RCL top, $\overline{L}_x(z)$ is small and nearly uniform; across the transition region it rises steadily; and in the LPL it levels off, with no proportional-to-$z$ growth anywhere in the measured range. Panel (d) reduces each case to two numbers — the average over the RCL and the value at the LPL midpoint. Inside the layer, the scale is nearly common to all wave-forced cases, 0.4--0.5$D$, and only $\approx 0.2D$ for the two pure currents. Above the layer, by contrast, the plateau level is not common but grows with the strength of the forcing: from 0.78$D$ (W1) to 1.22$D$ (W4) across the pure waves, from 0.57$D$ (C1) to 1.23$D$ (C2) for the currents, and 0.97--1.32$D$ across the wave-current cases.

\subsection{Summary of the kinematic characteristics of the oscillatory RCL}\label{sec:RCL thickness summary}

The observations of \S~\ref{sec:RCL top}--\S~\ref{sec:TPC} are summarised as follows. 

(1) The apparent RCL is thin in every flow examined, under a steady current as much as under waves: its top lies $0.1$--$0.2D$ above the crests, shows no trend with $A_{bm}/k_N$, agrees with the kinematic estimate~\eqref{eq:threshold height}, and preserves the ordering of the flow types over the threshold range $0.3$--$0.6$. 

(2) The thickness varies over the wave cycle in step with the near-bed velocity: the layer is thinnest shortly before each free-stream reversal, where the near-bed flow reverses, and thickest in the middle of the reinforced half-cycle, and a superimposed current raises the period-averaged top by about $0.1D$. 

(3) The integral length of the turbulence is 0.4--0.5$D$ inside the layer under wave forcing and about $0.2D$ under a pure current, while above the layer it increases to a larger value that varies from case to case (0.57--1.32$D$) and does not grow with $z$ within the measured range.

(4) The eddy length does not vary in step with the thickness: $L_x$ peaks in mid-deceleration, about $11^\circ$ before the RCL thickness minima.

\section{Dynamic processes inside the oscillatory RCL}\label{sec:dynamics}

This section examines the dynamics inside the RCL identified in \S~\ref{sec:RCL}. Two partitions are examined. \S~\ref{sec:TKE DKE} splits the kinetic energy between the turbulence ($k_t$) and the dispersive motion ($k_d$), and \S~\ref{sec:momentum transfer} splits the shear stress between the Reynolds and dispersive contributions.

\subsection{Energy partition inside the RCL}\label{sec:TKE DKE}

\begin{figure}
 \centering
  \includegraphics[width=\columnwidth]{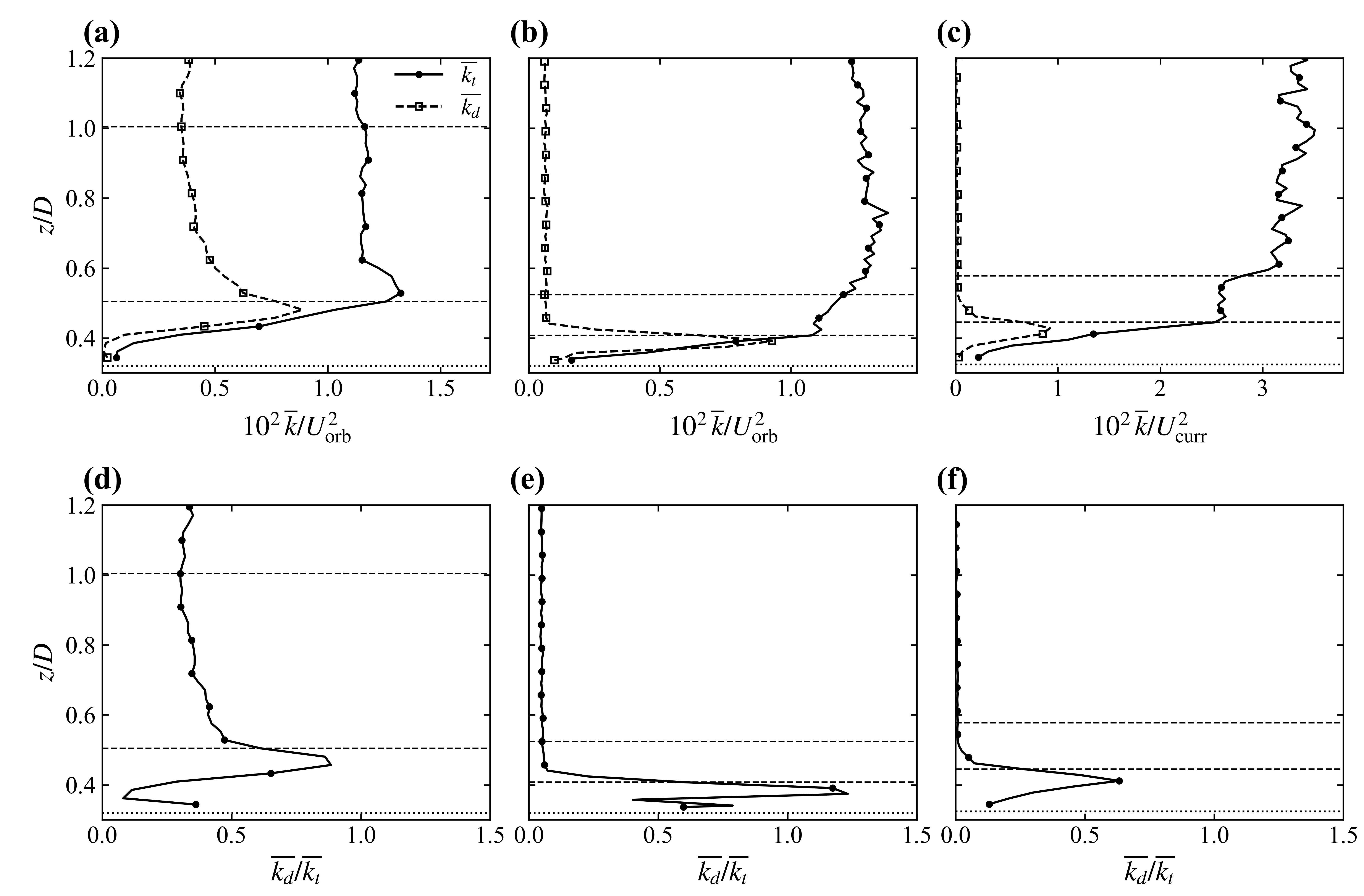}
  \caption{Time-averaged turbulent ($\overline{k_t}$, solid) and dispersive ($\overline{k_d}$, dashed) kinetic energy for the three exemplars, (a,d)~WC3, (b,e)~W3 and (c,f)~C2. (a--c)~Profiles scaled by $10^{2}$, note the per-panel abscissae. (d--f)~The ratio $\overline{k_d}/\overline{k_t}$ on a common abscissa. Horizontal dashed lines mark, in ascending order, the group-mean $z^{R}_{\mathrm{RCL}}$ and $z_{\mathrm{log,b}}$, the dotted line the crest.}
\label{Fig DKE TKE profile}
\end{figure}

\begin{figure}
 \centering
\includegraphics[width=\columnwidth]{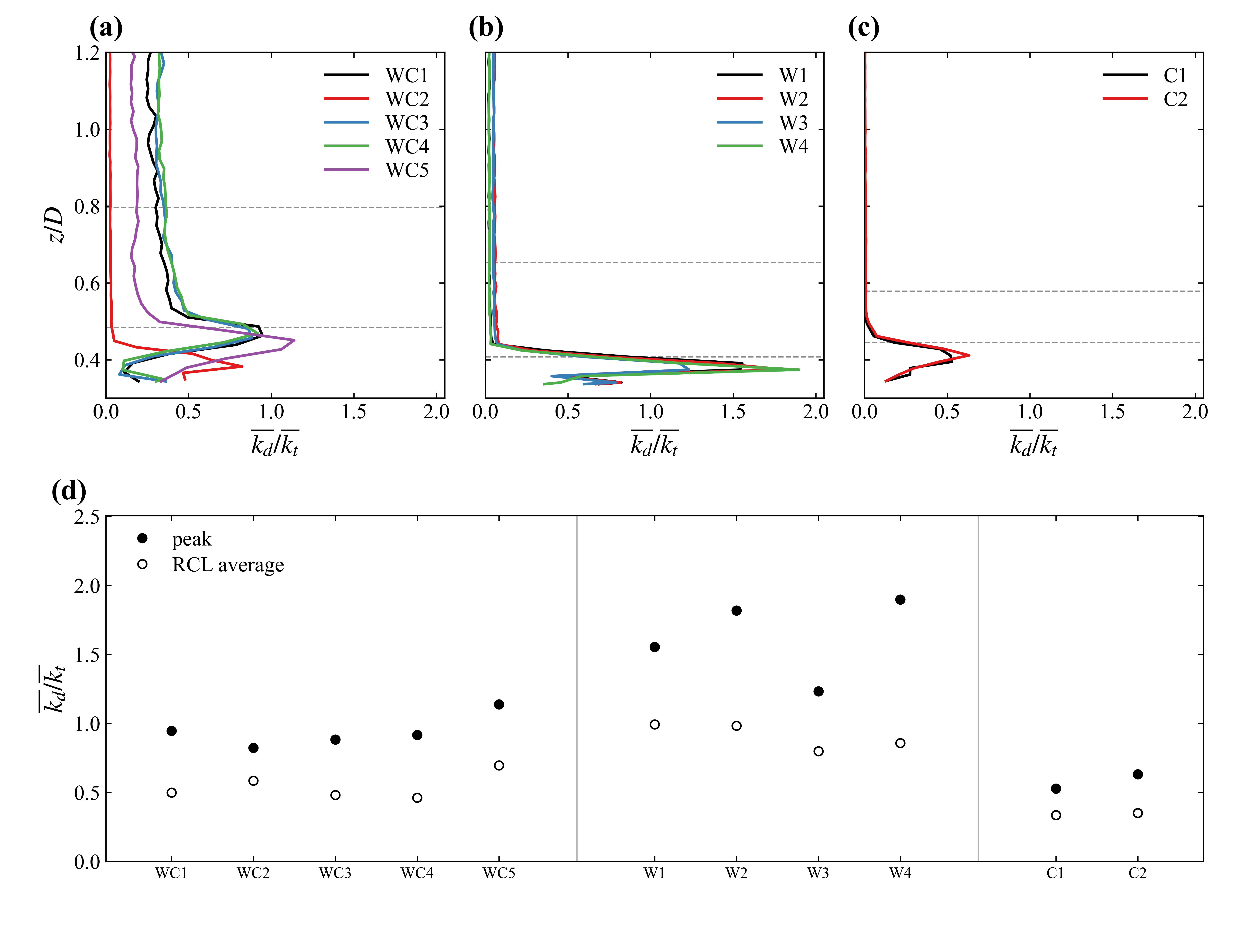}
  \caption{Energy partition for all eleven cases. (a--c)~Profiles of $\overline{k_d}/\overline{k_t}$ for the wave-current, pure wave and pure current groups. Grey dashed lines as in Fig.~\ref{Fig integral length summary}. (d)~Per-case summary. Vertical lines separate the groups.}
\label{Fig DKE TKE profile summary}
\end{figure}

Two kinetic energies are compared as time averages, since what matters here is their vertical structure rather than their phase modulation, which \S~\ref{sec:RCL} has already established for the dispersive field. Time-averaged profiles are also directly comparable across all three forcing types, including the steady currents. Fig.~\ref{Fig DKE TKE profile} shows the profiles for one case of each forcing type: wave-current WC3 (panels a,d), pure wave W3 (b,e), and pure current C2 (c,f). In panels (a--c), the turbulent kinetic energy $\overline{k_t}$ is small at the crest, rises through the RCL, and peaks at or just above the RCL top. The dispersive kinetic energy $\overline{k_d}$ instead peaks inside the layer, below the $\overline{k_t}$ maximum, where the inter-crest wakes reside, and drops sharply at the RCL top. The ratio panels (d--f) make the layering explicit: $\overline{k_d}/\overline{k_t}$ is maximal just below the layer top (about 0.9 for WC3, above unity for W3 and about 0.6 for C2) and falls off abruptly above the RCL in all three cases.

Fig.~\ref{Fig DKE TKE profile summary} extends the partition to all eleven cases, reducing each to two numbers: the peak ratio and the ratio averaged over the RCL. The peak is the maximum of the profile $\overline{k_d}(z)/\overline{k_t}(z)$, attained just below the RCL top in every case. The peak ratios are ordered by group: $1.2$--$1.9$ for the pure waves, $0.8$--$1.15$ for the wave-current cases, and $0.5$--$0.65$ for the pure currents. In the pure wave cases, the dispersive kinetic energy within the RCL thus reaches nearly twice the turbulent kinetic energy. The RCL-averaged ratios preserve the same ordering at lower values ($\approx 0.8$--$1.0$, $0.5$--$0.7$, and $\approx 0.35$).

Above the RCL, the groups separate. The pure wave and pure current ratios collapse to $\approx 0.05$. The wave-current cases show the same sharp drop at the RCL top, but retain a residual level of $\approx 0.15$--$0.35$ through the transition region and the LPL - the energy counterpart of the slowly decaying $\overline{R}(z)$ of table~\ref{tab:RCL top}.

\subsection{Momentum transport inside the RCL}\label{sec:momentum transfer}

\begin{figure}
 \centering
  \includegraphics[width=\columnwidth]{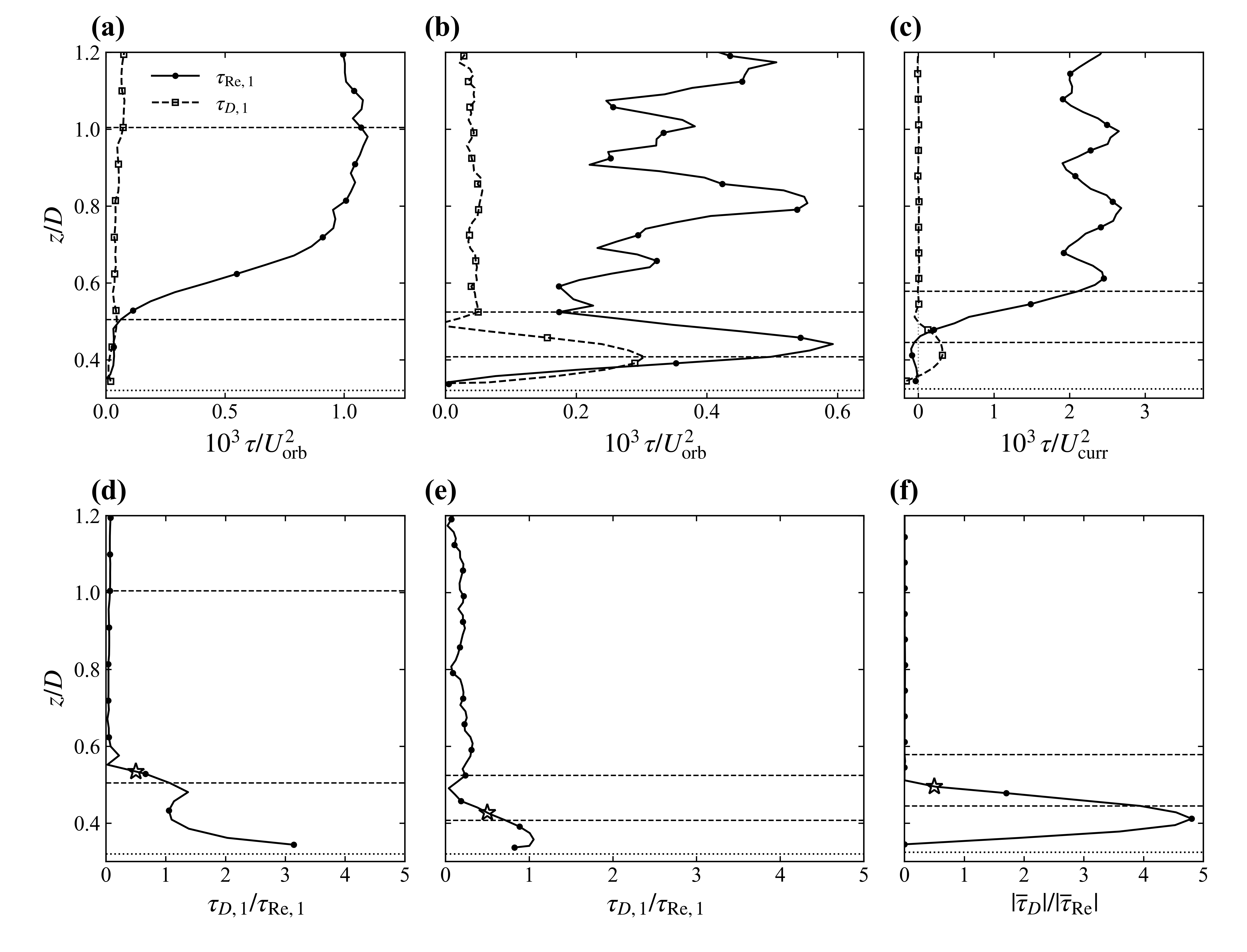}
  \caption{Reynolds (solid) and dispersive (dashed) stresses for (a,d)~WC3, (b,e)~W3 and (c,f)~C2. (a--c)~First-harmonic amplitudes for WC3 and W3, time-averaged stresses for C2 (dotted line marks zero), scaled by $10^{3}$. Note the per-panel abscissae. (d--f)~The stress ratio on a common abscissa. Stars mark $z_{0.5}$. Horizontal lines as in Fig.~\ref{Fig DKE TKE profile}.}
\label{Fig RS1 and DS1}
\end{figure}

\begin{figure}
 \centering
  \includegraphics[width=\columnwidth]{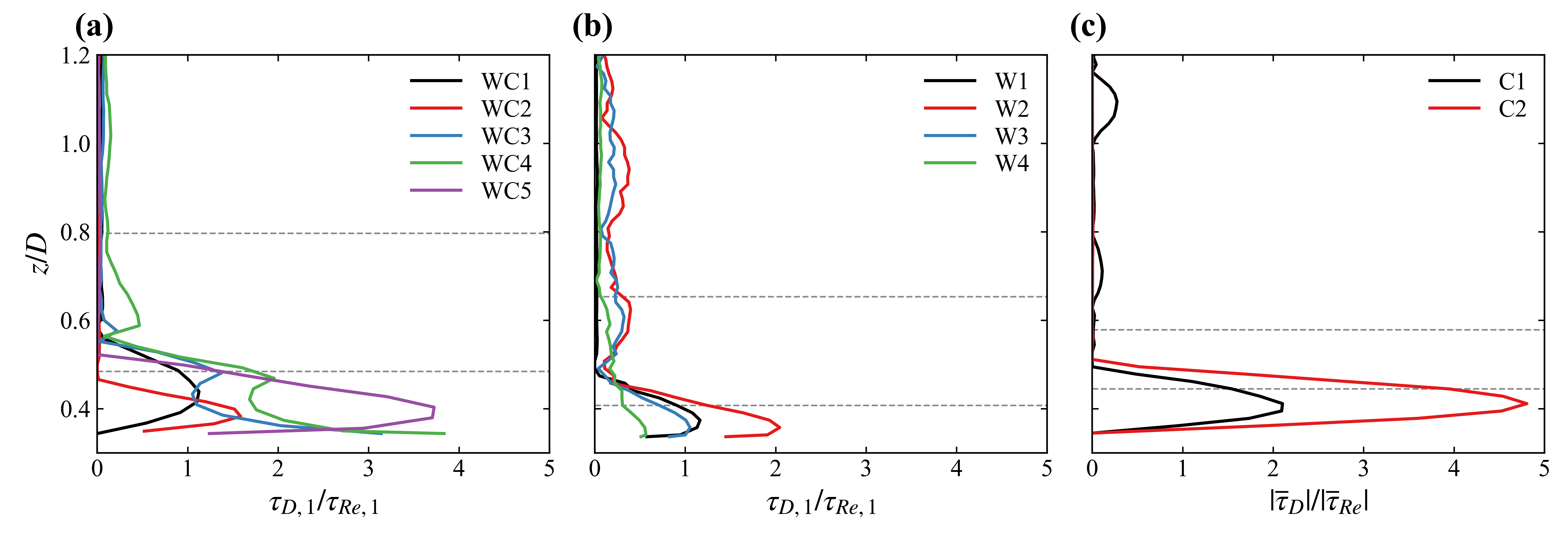}
  \caption{Dispersive-to-Reynolds stress ratio for all eleven cases: profiles for the wave-current, pure wave and pure current groups, ratios defined as in Fig.~\ref{Fig RS1 and DS1}. Grey dashed lines as in Fig.~\ref{Fig integral length summary}.}
\label{Fig RS1 and DS1 summary}
\end{figure}

\begin{figure}
 \centering
  \includegraphics[height=8cm]{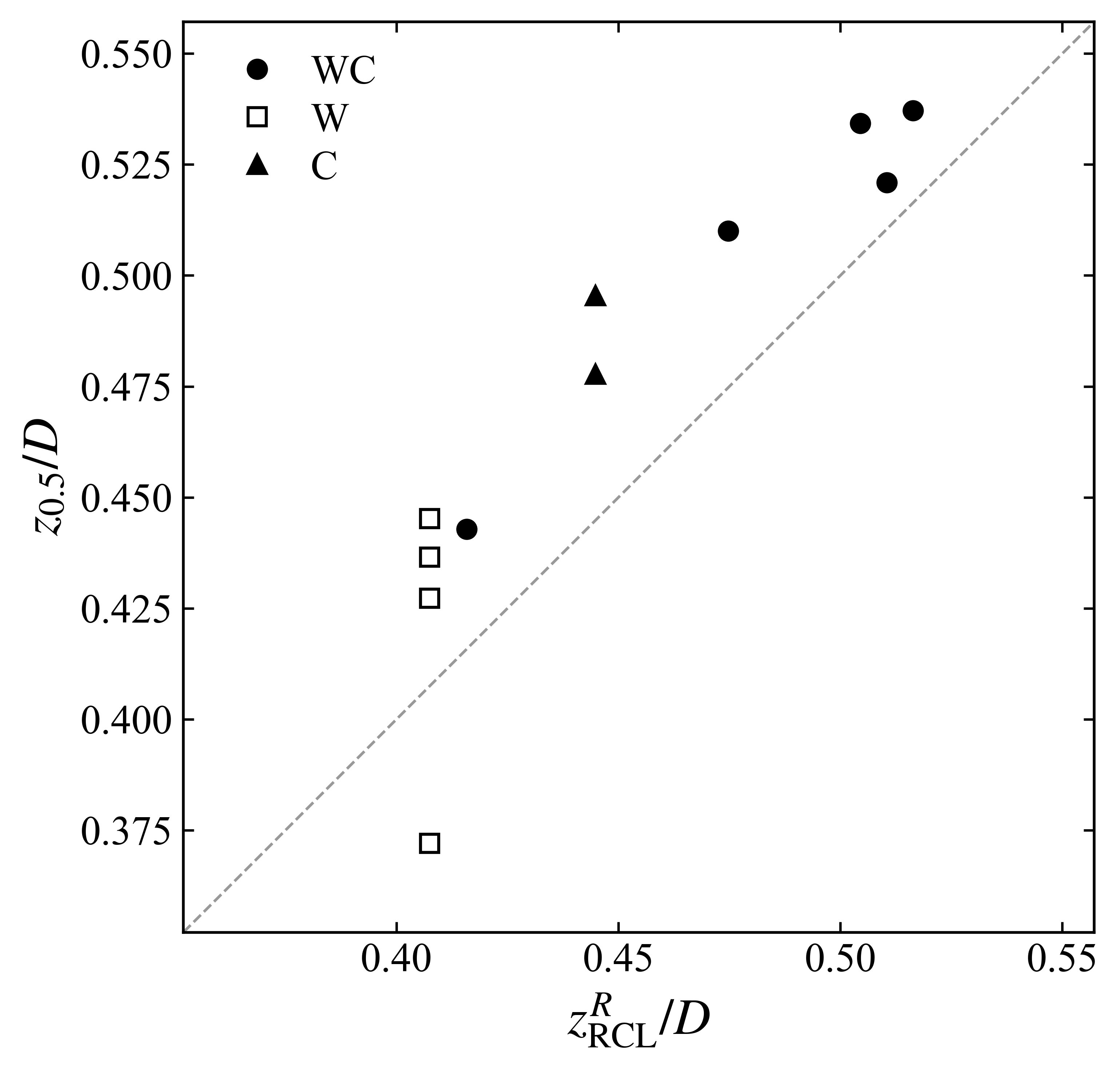}
  \caption{Cross-validation of the RCL top: the stress-based $z_{0.5}$ against the velocity-based $z^{R}_{\mathrm{RCL}}$ for the wave-current (filled circles), pure wave (open squares) and pure current (filled triangles) groups. The dashed line is the identity.}
\label{Fig stress crossval}
\end{figure}

Fig.~\ref{Fig RS1 and DS1} shows the corresponding partition of the momentum flux: the first-harmonic amplitudes of the Reynolds and dispersive shear stresses of \eqref{eq:RS and DS}, $\tau_{\mathrm{Re},1}$ and $\tau_{D,1}$ (time-averaged stresses for the pure current), for the same three cases. In panels (a--c), the Reynolds stress is small at the crest and rises through the RCL to its outer level, while the dispersive stress is confined to the layer: it peaks inside the RCL and collapses at its top. The ratio panels (d--f) quantify the layering: near the crest, the dispersive stress reaches three times $\tau_{\mathrm{Re},1}$ in the wave-current case and nearly five times the time-averaged Reynolds stress in the pure current, before the ratio falls below 0.5 (stars) within a fraction of a diameter. For the pure current case, the ratio is formed from stress magnitudes.

Fig.~\ref{Fig RS1 and DS1 summary} extends the partition to all eleven cases. The profiles repeat one shape: near the crest, the dispersive stress is comparable to the Reynolds stress and dominant in most cases, with peak ratios of roughly 1.1--3.8 for the wave-current group, 0.6--2 for the pure waves, and 2--5 for the pure currents, and it collapses above the RCL.

The stress partition defines its own layer top, and Fig.~\ref{Fig stress crossval} compares it with the velocity-based one. The stress-based marker $z_{0.5}$ is the first height above the crests at which the dispersive-to-Reynolds stress ratio falls below 0.5, in direct analogy with the detection of $z^R_{\mathrm{RCL}}$ from the velocity. Plotted against each other, the two markers cluster along the identity for all eleven cases, agreeing to within $0.05D$, that is, within three measurement points. The offsets are not random in sign: the wave-current and pure current points sit slightly above the identity, by $0.01$--$0.05D$, while the pure wave points scatter about it. The two detections rest on independent quantities: $z^R_{\mathrm{RCL}}$ measures how high the velocity deviation remains large, and $z_{0.5}$ measures how high the wakes carry organised momentum. Their coincidence establishes that the dynamical RCL and the apparent RCL are the same layer.

These stresses must be read against the resolving limits of the PIV. The finite interrogation window low-pass filters the velocity field and attenuates the measured Reynolds stress towards the bed — to about $75\%$ of the momentum-integral estimate at $z \approx 5$ mm for the sinusoidal wave over this bed\cite{Yuan2014} — and the attenuation is strongest where the eddies are finest, inside the layer. The dispersive stress, resolved at the marble scale, is much less affected. The measured near-crest Reynolds stress is therefore an underestimate, and the stress ratios of Figs.~\ref{Fig RS1 and DS1} and~\ref{Fig RS1 and DS1 summary} within the layer are correspondingly overestimates.

\subsection{Summary of the dynamic characteristics of the oscillatory RCL}\label{sec:dynamics summary}

The observations of \S~\ref{sec:TKE DKE}--\S~\ref{sec:momentum transfer} are summarised as follows. 

(1) Within the RCL, the dispersive kinetic energy is of the same order as the turbulent kinetic energy, with peak ratios ordering as pure current (0.5--0.65) < wave-current (0.8--1.15) < pure wave (1.2--1.9). 

(2) Within the RCL, the dispersive stress matches, and near the crest exceeds, the Reynolds stress, by up to a factor of 3--5. Since attached flow carries no dispersive stress \eqref{eq:zero dispersive stress}, separated wakes are present and transport organised momentum within the layer. 

(3) The stress-based top $z_{0.5}$ coincides with the velocity-based top $z^R_{\mathrm{RCL}}$ to within $0.05D$ in every case: the dynamical RCL and the apparent RCL are the same layer. These observations are interpreted in \S~\ref{sec:discussion}.

\section{Discussion}\label{sec:discussion}

\S~\ref{sec:RCL} and \S~\ref{sec:dynamics} are observational: they establish a thin apparent RCL whose thickness and internal eddy scale vary over the wave cycle on different schedules, and a dynamical layer, coincident with it, in which the dispersive motion carries a share of the energy and momentum comparable to the turbulence. This section discusses these observations together. \S~\ref{sec:Wake-Evolution Framework} organises the phase behaviour into the life cycle of the RCL, and \S~\ref{sec:why thin} examines what controls its thickness.

\subsection{The life cycle of the RCL}\label{sec:Wake-Evolution Framework}

\begin{figure}
 \centering
  \includegraphics[width=\columnwidth]{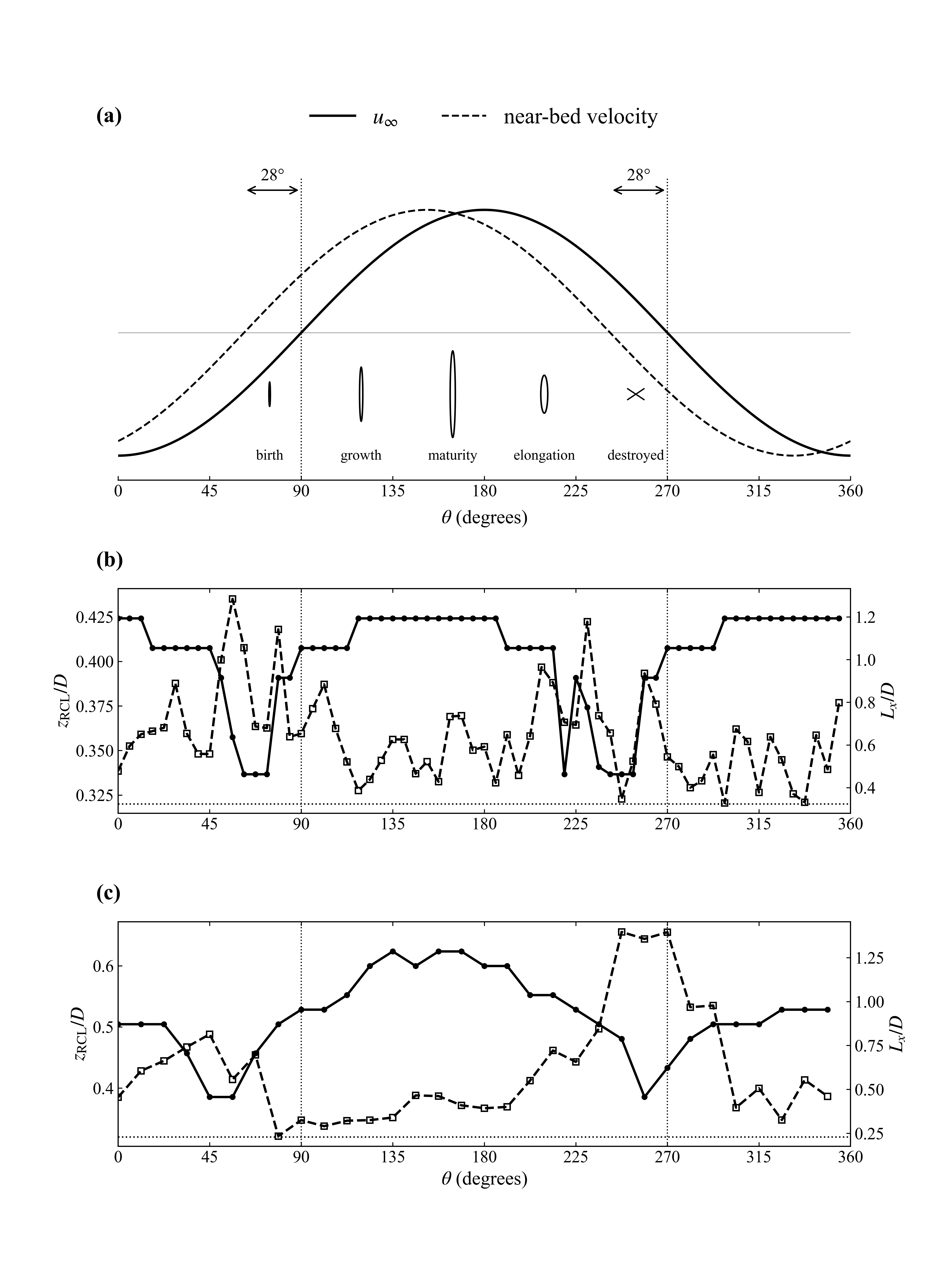}
  \caption{Wake life cycle. (a)~Schematic: free-stream and near-bed velocity, the latter with the $28^\circ$ lead for W3, and the wake stages between successive near-bed reversals. (b,c)~Phase-averaged RCL top $z_{\mathrm{RCL}}(\theta)/D$ (solid, left axis) and integral length at that height, $L_x/D$ (dashed, right axis), for (b)~W3 and (c)~WC3. Vertical dotted lines mark the free-stream reversals, the grey line in (a) zero velocity, and the horizontal dotted line in (b,c) the crest.}
\label{Fig wake conceptual}
\end{figure}

The apparent RCL of \S~\ref{sec:RCL} is not a permanent structure: it is destroyed and rebuilt twice per wave period. Fig.~\ref{Fig wake conceptual} panel (a) sketches this life cycle, and panels (b,c) show its measured signature for the pure wave W3 and the wave-current case WC3. The near-bed velocity sets the timing throughout, leading the free stream by roughly $28^\circ$ - the lead of the thickness minimum over the free-stream reversal in W3 (panel b), which the quasi-steady response of \S~\ref{sec:kinematic decay} identifies with the near-bed lead. The layer is born just after the near-bed flow reverses: a new wake forms in the gap behind each crest, the layer is at its thinnest, and the structures inside it are the finest of the whole cycle ($L_x \approx 0.2$--$0.3D$). During the acceleration, the layer grows. The oscillatory pressure gradient acts at every elevation, inside the inter-crest cavities as much as above them, so the interstitial flow responds immediately and the wakes strengthen with the near-bed velocity. The layer top rises while $L_x$ stays low, the young wakes gaining strength but not length.

The layer matures when the near-bed velocity peaks: in Fig.~\ref{Fig wake conceptual} (c), the top is highest at $135$--$170^\circ$, around the near-bed maximum near $152^\circ$ rather than the free-stream maximum at $180^\circ$. During the deceleration, the layer thins while its structures stretch. The weakening near-bed flow lowers the deviation amplitude, and the top falls with it, while the adverse pressure gradient elongates the separated structures and drives $L_x$ to its cycle maximum ($1.40D$ at $245$--$268^\circ$ in WC3). The layer is destroyed when the near-bed flow reverses, ahead of the free stream. The phase-locked deviation collapses to the thickness minimum, and the elongated structures detach from the crests and are absorbed into the background turbulence. The $L_x$ maximum precedes the thickness minimum by about $11^\circ$; the data establish the timing, not that one causes the other. The next half-cycle repeats the sequence in the opposite direction.

Seen this way, the phase behaviour of \S~\ref{sec:RCL} has a simple structure: the thickness and the eddy length are two properties of the same wake field that answer to different clocks. The thickness answers to the instantaneous forcing. Its vertical decay rate is frozen by the packing (\S~\ref{sec:kinematic decay}), so the only quantity free to vary is the amplitude of the deviation, and the amplitude follows the near-bed velocity quasi-steadily because the wakes respond fast: the eddy turnover time $\tau_e \sim D/u_*$ is 0.08--0.26 s against periods of 6.25 and 12.5 s, so $\tau_e/T \approx 0.01$--$0.04$. This is why the layer is thickest at the near-bed velocity maximum, why its minima lead the free-stream reversals (\S~\ref{sec:RCL top}), and why large swings in forcing move the top only modestly: the forcing amplitude enters~\eqref{eq:threshold height} only through $R_c$, inside the logarithm, so a doubling of $R_c$ raises the top by just $\ell\ln 2$. The eddy length instead answers to the history of the forcing: it measures how far the structures have developed since birth, so it grows through the deceleration and peaks in the elongation stage. Hence the two quantities reach their extrema at different phases, and the eddy length leads.

\subsection{Controls on the RCL thickness}\label{sec:why thin}

The measured layer is an order of magnitude thinner than the steady-flow estimate for the same geometry, and the kinematics of \S~\ref{sec:kinematic decay} gives the first-order reason. Above a periodic bed, the deviation field cannot decay more slowly than $\mathrm{e}^{-z/\ell}$ with $\ell = s/2\pi$, so close packing ($s \approx D$) caps any velocity-based layer at $\ell \ln(R_c/R_{\mathrm{thr}})$, a fraction of a diameter, and this is what is measured in \S~\ref{sec:RCL top}. The only known mechanism that carries the roughness influence higher, the collective canopy instability, requires a permeable interior: drag distributed over a depth and an inflectional mean profile at its top \cite{raupach1996coherent, finnigan2000}. A single layer of spheres on a wall has neither, and its vorticity is confined to the crest region, which is exactly the condition under which the kinematic decay governs the field above. The steady-current cases return the same thin layer as the waves ($0.125D$ against $0.09D$), so the thinness owes nothing to the oscillation: it is a property of the bed.

Within that thin range, the ordering of the forcing types — waves thinnest, currents thicker, combined flows thickest — must enter through the one free parameter of \eqref{eq:threshold height}, the crest-level amplitude $R_c$, and $R_c$ reflects how the flow in the inter-crest cavities is driven. A wave drives the cavities directly: the oscillatory pressure gradient, $-\rho^{-1}\partial p_w/\partial x = \partial u_\infty/\partial t$, acts at every elevation, so the interstitial flow moves with the flow above the crests and the crest-level deviation is comparatively small. A current cannot: its streamwise pressure gradient is negligible, momentum reaches the cavities only as turbulent shear from above, and the form drag absorbs much of that flux at the crests, leaving a sheltered cavity flow, a deeper crest-level velocity deficit, and a larger $R_c$. A superimposed current drives the wake field hardest of all: it holds the near-bed flow in one direction through the reinforced half-cycle, so the wakes are driven harder and for longer, and $R_c$ is the largest of the three groups, setting the highest RCL top. The same bias raises the period-averaged top by $\approx 0.1D$ and produces the unequal $L_x$ lobes of \S~\ref{sec:RCL}. The measured crest amplitudes follow this ordering in three bands ($R_c \approx 0.96$--$1.06$ for the pure waves, $\approx 1.2$ for the pure currents, and $1.31$--$1.62$ for the wave-current cases). The bands set a ceiling rather than the top itself: the ordering holds between the forcing types, not within them, and the wave-current tops in particular do not order by $R_c$.

Thinness does not mean that the roughness influence is weak. The same close packing that minimises the spacing forces every near-crest streamline to separate over a discrete crest, and \S~\ref{sec:dynamics} shows that the resulting wakes carry a dispersive stress that matches and locally exceeds the Reynolds stress. The thickness and the dynamical strength of the layer are set by different aspects of the same geometry.

Three factors could make the layer thicker, and they differ greatly in strength. The first is the element spacing: $\ell = s/2\pi$ grows linearly with $s$, so doubling the spacing doubles the layer. The second is permeability: a bed the flow can penetrate distributes the drag over a depth and produces an inflected mean profile, which activates the canopy instability and its layer of two to five element heights. The third, and weakest, is the strength of the wakes, raised by the forcing or by sharper elements, which enters only logarithmically through $R_c$ in \eqref{eq:threshold height}. Varying the spacing and the permeability separately would test the first two factors and locate where roughness behaviour ends, and canopy behaviour begins.

\section{Conclusions}\label{sec:conclusions}

This study set out to resolve an apparent contradiction. Oscillatory boundary layer experiments over a densely packed marble bed, at gravel-scale relative roughness $A_{bm}/k_N = O(10$--$100)$, recover accurate logarithmic profiles within millimetres of the roughness crests\cite{Yuan2014, Yuan2015}. Yet steady-flow theory places a roughness-controlled layer of two to five element heights over the same region\cite{Raupach1991, Jimenez2004}. A re-analysis of the raw PIV records with a triple decomposition, separating the coherent motion locked to the marbles from the stochastic turbulence resolves it: the RCL over such a bed is thin. Across eleven flow conditions, its top lies only $0.1$--$0.2D$ above the crests, with no trend in $A_{bm}/k_N$. This is the scale the packing imposes kinematically: above a periodic bed, the deviation field decays over $\ell = s/2\pi$, so close packing caps any velocity-based layer at a fraction of a diameter, an order of magnitude below the canopy estimate. Above the layer, a clean logarithmic profile layer occupies the bulk of the wave boundary layer.

Under oscillatory forcing, the layer is a periodic structure with a life cycle: born as the near-bed flow re-establishes after each reversal, grown through the acceleration, thickest in the vicinity of the near-bed velocity maximum, thinned and stretched through the deceleration, and destroyed as the near-bed flow reverses ahead of the free stream. The thickness follows the instantaneous near-bed velocity quasi-steadily ($\tau_e/T \approx 0.01$--$0.04$), whereas the streamwise eddy length follows the history of the forcing, peaking in the deceleration about $11^\circ$ before the thickness minima. Inside the layer, the eddies are locked to the bed geometry, 0.4--0.5$D$ under wave forcing and $0.2D$ under a pure current, while above it they settle onto a forcing-dependent plateau of 0.6--1.32$D$ with no $\kappa z$ growth in the measured range. A superimposed current biases the cycle through the reinforced half-cycle, raising the period-averaged top by about $0.1D$ and extending the roughness influence farther above the layer than either forcing alone.

The thin layer is dynamically active. Within it, the dispersive kinetic energy rivals the turbulent kinetic energy, reaching nearly twice it under pure wave forcing, and the dispersive stress matches, and near the crest exceeds, the Reynolds stress before collapsing at the layer top. Because attached flow over a fore--aft symmetric element carries no dispersive stress, these magnitudes establish that separated wakes transport organised momentum within the layer. The stress-based layer top coincides with the velocity-based one to within $0.05D$ in every case, so the dynamical and the apparent RCL are the same layer, and a substantial share of the near-crest stress reported as Reynolds stress in earlier measurements over this bed\cite{Yuan2014, Yuan2015} is in fact dispersive: the momentum flux carried by the element-locked wakes.

The condition under which the logarithmic layer exists in this regime follows. The RCL together with the transition region must remain small against $\delta_w$, the layer top itself being $\ell \ln(R_c/R_{\mathrm{thr}})$, and here the two differ by more than an order of magnitude. The margin is expected to be controlled by the bed rather than by the forcing: the layer should thicken linearly with the element spacing; permeability would activate the canopy mechanism and could push the layer to two to five element heights, comparable to $\delta_w$ itself; the forcing and the element shape enter only logarithmically. These expectations remain to be tested, and sparser, permeable, and mobile beds, where the margin narrows, are the natural next step.

\begin{acknowledgments}
This research is funded by China’s MST Key R\&D (Grant 2023YFC3206201).
\end{acknowledgments}

\section*{Data Availability Statement}

The data that support the findings of this study are available from the corresponding author upon reasonable request.

\bibliography{aipsamp}

\end{document}